\documentclass[a4paper,11pt]{article}
\usepackage{jcappub} 

\usepackage{graphicx}
\usepackage[varg]{txfonts}
\usepackage{graphicx}
\usepackage{psfrag}
\usepackage{hyperref}
\usepackage{xcolor}
\usepackage{tabularx}
\usepackage{lipsum}
\usepackage[normalem]{ulem}
\usepackage{amsmath}
\usepackage[labelfont=bf, labelsep=period]{caption}
\usepackage{subcaption}
\usepackage{mathtools}
\usepackage{natbib}
\usepackage{balance}
\usepackage{adjustbox}
\usepackage{float} 
\usepackage{placeins}
\usepackage{caption}
\usepackage{stfloats}
\usepackage{pifont}
\usepackage{xcolor}
\usepackage{dsfont}
\usepackage{multirow}
\usepackage{changepage}
\usepackage{makecell}
\usepackage{soul}

\usepackage[colorinlistoftodos]{todonotes}
\usepackage{easyReview}

\usepackage{enumitem}

\usepackage{siunitx} 
\usepackage[switch,displaymath,mathlines]{lineno}
\usepackage{cleveref}

\title{Requirements on the gain calibration for \textit{LiteBIRD} polarisation data with blind component separation}

\def\reff@jnl#1{{\rm#1\/}}

\def\aj{\reff@jnl{AJ}}                  
\def\araa{\reff@jnl{ARA\&A}}            
\def\apj{\reff@jnl{ApJ}}                
\def\apjl{\reff@jnl{ApJ}}               
\def\apjs{\reff@jnl{ApJS}}              
\def\ao{\reff@jnl{Appl.Optics}}         
\def\apss{\reff@jnl{Ap\&SS}}            
\def\aap{\reff@jnl{A\&A}}               
\def\aapr{\reff@jnl{A\&A~Rev.}}         
\def\aaps{\reff@jnl{A\&AS}}             
\def\azh{\reff@jnl{AZh}}                        
\def\baas{\reff@jnl{BAAS}}              
\def\jcap{\reff@jnl{JCAP}}              
\def\jltp{\reff@jnl{J.~Low~Temp.~Phys.}}
\def\jrasc{\reff@jnl{JRASC}}            
\def\memras{\reff@jnl{MmRAS}}           
\def\mnras{\reff@jnl{MNRAS}}            
\def\pra{\reff@jnl{Phys.~Rev.~A}}       
\def\prb{\reff@jnl{Phys.~Rev.~B}}       
\def\prc{\reff@jnl{Phys.~Rev.~C}}       
\def\prd{\reff@jnl{Phys.~Rev.~D}}       
\def\prl{\reff@jnl{Phys.~Rev.~Lett}}    
\def\physrep{\reff@jnl{Phys.Rep.}}      
\def\ptep{\reff@jnl{PTEP}}              
\def\pasp{\reff@jnl{PASP}}              
\def\pasj{\reff@jnl{PASJ}}              
\def\qjras{\reff@jnl{QJRAS}}            
\def\skytel{\reff@jnl{S\&T}}            
\def\solphys{\reff@jnl{Solar~Phys.}}    
\def\sovast{\reff@jnl{Soviet~Ast.}}     
 \def\ssr{\reff@jnl{Space~Sci.Rev.}}    
\def\zap{\reff@jnl{ZAp}}                
\def\nat{\reff@jnl{Nature}}             

\author[1,2,3]{F.\,Carralot,}
\author[1,3,4]{A.\,Carones,}
\author[1,3,5]{N.\,Krachmalnicoff,}
\author[6,7]{T.\,Ghigna,}
\author[8]{A.\,Novelli,}
\author[9,10,11]{L.\,Pagano,}
\author[8,12]{F.\,Piacentini,}
\author[1,3,5]{C.\,Baccigalupi,}
\author[13]{D.\,Adak,}
\author[4]{A.\,Anand,}
% ...
\affiliation[1]{International School for Advanced Studies (SISSA), Via Bonomea 265, 34136, Trieste, Italy}
\affiliation[2]{Università di Trento, Dipartimento di Fisica, Via Sommarive 14, 38123, Trento, Italy}
\affiliation[3]{INFN Sezione di Trieste, via Valerio 2, 34127 Trieste, Italy}
\affiliation[4]{Dipartimento di Fisica, Università di Roma Tor Vergata, Via della Ricerca Scientifica, 1, 00133, Roma, Italy}
\affiliation[5]{IFPU, Via Beirut, 2, 34151 Grignano, Trieste, Italy}
\affiliation[6]{International Center for Quantum-field Measurement Systems for Studies of the Universe and Particles (QUP), High Energy Accelerator Research Organization (KEK), Tsukuba, Ibaraki 305-0801, Japan}
\affiliation[7]{Kavli Institute for the Physics and Mathematics of the Universe (Kavli IPMU, WPI), UTIAS, The University of Tokyo, Kashiwa, Chiba 277-8583, Japan}
\affiliation[8]{Dipartimento di Fisica, Università La Sapienza, P. le A. Moro 2, Roma, Italy}
\affiliation[9]{Dipartimento di Fisica e Scienze della Terra, Università di Ferrara, Via Saragat 1, 44122 Ferrara, Italy}
\affiliation[10]{INFN Sezione di Ferrara, Via Saragat 1, 44122 Ferrara, Italy}
\affiliation[11]{Université Paris-Saclay, CNRS, Institut d’Astrophysique Spatiale, 91405, Orsay, France}
\affiliation[12]{INFN Sezione di Roma, P.le A. Moro 2, 00185 Roma, Italy}
\affiliation[13]{Instituto de Astrofísica de Canarias, E-38200 La Laguna, Tenerife, Canary Islands, Spain}
% ...

\emailAdd{fcarralo@sissa.it}
\emailAdd{acarones@sissa.it}
\emailAdd{nkrach@sissa.it}

\abstract{
The detection of primordial $B$ modes of the cosmic microwave background (CMB) could provide information about the early stages of the Universe's evolution. The faintness of this signal requires exquisite calibration accuracy and control of instrumental systematic effects which otherwise could bias the measurements. In this work, we study the impact of an imperfect relative polarisation gain calibration on the recovered value of the tensor-to-scalar ratio $r$ for the \textit{LiteBIRD} experiment, through the application of the blind Needlet Internal Linear Combination (NILC) foreground-cleaning method. We derive requirements on the relative calibration accuracy of the overall polarisation gain ($\Delta g_\nu$) for each \textit{LiteBIRD} frequency channel. 
Our results show that minimum variance techniques, as NILC, are less sensitive to systematic gain calibration uncertainties compared to a parametric approach, if the latter is not equipped with a proper modelling of these instrumental effects. In this study, the most stringent requirements are found in the channels where the CMB signal is relatively brighter, with the tightest constraints at 166\,GHz ($\Delta {g}_\nu \approx 0.16 \%$). This differs from the outcome of an analogous analysis performed with a parametric method, where the tightest requirements are obtained for the foreground-dominated channels. Gain calibration uncertainties, corresponding to the derived requirements, are then simultaneously propagated into all frequency channels. By doing so, we find that the overall impact on estimated $r$ is lower than the total gain systematic budget for \textit{LiteBIRD} approximately by a factor $5$, due to the correlations of the impacts of gain calibration uncertainties in different frequency channels.
In order to decouple the systematic effect from the specific choice of the model, we derive the requirements assuming constant spectral parameters for the foreground emission. To assess the robustness of the obtained results against more realistic scenarios, we repeat the analysis assuming sky models of intermediate and high complexity.
In these further cases, we adopt an optimised NILC pipeline, called the Multi-Clustering NILC (MC-NILC). We find that the impact of gain calibration uncertainties on $r$ is lower than the \textit{LiteBIRD} gain systematics budget for the intermediate-complexity sky model. For the high-complexity case, instead, it would be necessary to tighten the requirements by a factor $1.8$.} 

\keywords{CMBR polarisation --  CMBR experiments}

\begin{document}
\maketitle

\section{Introduction} \label{int}

 Measurements of temperature anisotropies of the Cosmic Microwave Background (CMB) \citep{CMB} by the \textit{COBE} \cite{COBE}, \textit{WMAP} \cite{WMAP} and \textit{Planck} \cite{Planck2013,Planck2015,Planck2018} spacecraft and BOOMERanG \cite{boomerang}, SPT \cite{Balkenhol_2023} and ACT \cite{madhavacheril2023atacama} among other sub-orbital experiments\footnote{\url{https://lambda.gsfc.nasa.gov/}} led to major advancements in cosmology, allowing us to precisely constrain cosmological parameters in models capable of describing the evolution of the Universe. 
 In the last decade, efforts have been focused on the analysis of the CMB polarisation signals, which could serve as an additional observational window into the early Universe. Indeed, the standard model of Cosmology predicts that the Universe experienced a phase of exponential expansion perhaps $ 10^{-36}-10^{-34} $s after the Big Bang, named \emph{cosmic inflation} \cite{inf}. Such expansion magnified quantum fluctuations to cosmological scales and also generated tensor perturbations that produce a specific signature in the CMB polarisation: the $B$-mode pattern \cite{1997kamion,HuWhite,SeljakZaldarriaga}. Therefore, the detection of primordial CMB $B$ modes could enable the estimation of the amplitude of primordial gravitational waves quantified by the tensor-to-scalar ratio parameter $r$ \cite{infplanck} and potentially confirm the inflationary scenario. Due to the expected amplitude of such a signal (at least $1000$ times weaker than temperature anisotropies), primordial $B$ modes have not been detected yet, with current upper bounds on the tensor-to-scalar ratio $r\lesssim 0.03\ (95\%\ \text{CL})$ \cite{tristram2022,galloni2023}. Their detection thus represents one of the main goals of future CMB missions. In order to probe the inflationary paradigm, it is essential to target large angular scales by measuring two main power spectrum features: the reionisation bump ($\ell \lesssim 10$),
 associated with the scattering of CMB photons with free electrons released during cosmic reionisation, and the recombination bump ($\ell \sim 80$) \cite{polarbear,bicep}, which instead corresponds to the imprint of primordial tensor perturbations at the recombination epoch. 
On smaller angular scales, CMB $B$ modes are also generated from lensed CMB polarisation $E$ modes due to the gravitational interaction of CMB photons with the intervening cosmic large scale structure \cite{Lensing,Lensing2}. Such a signal has already been measured by    SPTpol \cite{SPTpol}, ACTpol \cite{ACTPOL}, PolarBear \cite{polarbear} and BICEP2/Keck \cite{bicep} ground based experiments.

Despite huge progress in terms of instrumental sensitivity, detecting primordial $B$ modes still remains extremely challenging. 
One of the major impediments for an accurate measurement of the CMB polarisation signal is the contamination due to Galactic emission \cite{Fg}. Physical processes occurring within our Galaxy induce complex emission, that is challenging to model and that must be subtracted for any scientific exploitation of CMB data.  In the framework of CMB polarisation analysis, we can safely neglect some Galactic microwave radiative processes like free--free radiation \cite{Rybicki,2011MNRAS.418..888M}, anomalous microwave emission (AME) \cite{2011ame,AME_commander,2017MNRAS.464.4107G,ame_quijote} or CO molecular lines \cite{CO}, since they are characterised by a very low polarisation fraction ($\lesssim 1\%$), and hence we need to consider only synchrotron and thermal dust polarised emission \cite{dustplanck,dustcomp}. The synchrotron radiation, that is dominant at low frequencies ($\lesssim70\,$GHz), is generated by cosmic-ray electrons that are accelerated by the Galactic magnetic field. At high frequencies ($ \gtrsim 100\,$GHz), aspherical dust grains in the interstellar medium are heated by stellar ultraviolet radiation and re-emit far-infrared radiation with a polarisation fraction close to 20$\%$. These two emission mechanisms are obviously prominent around the Galactic plane, but are also clearly detectable at higher latitudes \cite{Skalidis2018}. In the context of the \textit{Planck} mission, various foreground cleaning procedures have been employed \cite{compsepplanck}. Among them, we mention two categories: (i) parametric-fitting \citep{Commander, param, Bsecret,Azzoni2021,Vacher2022,Commander3}, which recovers the CMB signal by marginalising over the spectral parameters of Galactic foregrounds; and (ii) the so-called `blind' methods \cite{ILCWMAP,ILCharmonic,genILC,constrainedILC,NILC_dela,MCNILC}, whose purpose is to recover a cleaned CMB blackbody signal, without any assumption on the spectral energy distribution (SED) of foreground emission. Methods of the latter class, in most cases, are also referred to as \emph{minimum-variance} techniques, since they reconstruct the CMB signal as the minimum variance solution from the linear combination of multi-frequency observations, thus maximally reducing the foreground contamination in the $2$-point statistics \cite{Tegmark_2003}.

The second main source of uncertainty is the presence of instrumental systematic effects, arising from non-idealities and imperfect characterisation of the instrument. As mentioned above, the weak amplitude of the $B$-mode signal requires an exquisite degree of control of both polarised Galactic emission and systematic effects. Therefore, it is essential to develop techniques that are able to handle multiple sources of uncertainty that could lead to biases in the reconstructed primordial $B$ modes.

Among future CMB experiments, \textit{LiteBIRD} \cite{LiteBIRD,PTEP,spie2024} (Lite (Light) satellite for the studies of $B$-mode polarisation and Inflation from cosmic background Radiation Detection), is a space-borne experiment selected by the Japanese Aerospace Exploration Agency (JAXA) and  which is currently backed by a world-wide collaboration. \textit{LiteBIRD} will perform $3$ years of full-sky observations to target an overall uncertainty on the tensor-to-scalar ratio of $\sigma_r \leq 10^{-3}$ by measuring both the recombination and reionisation bumps. The scope of this paper is to study the impact of imperfect photometric gain calibration on the estimate of the tensor-to-scalar ratio for the \textit{LiteBIRD} experiment. Therefore, in this paper, we set requirements on the gain calibration accuracy for each of \textit{LiteBIRD}’s frequency channels, which will allow it to fulfill the budget allocated to gain systematics, $\Delta r= 6.5 \times {10}^{-6}$ \cite{PTEP}. In this work, we refer to the relative polarisation gain as the 
gain calibration of polarisation data (\textit{Q} and \textit{U} Stokes parameter maps) relative to a specific frequency channel. In practice, this corresponds to the product of the intensity calibration and polarisation efficiency. The absolute gain, which is associated with a common amplitude factor affecting all frequency channels, does not impact the component-separation outcome and therefore we do not consider it in our analysis.

In this work, we make use of Needlet Internal Linear Combination (NILC) \cite{NILC,NILC2}, a blind component-separation technique that performs a linear combination of frequency maps in needlet space, in order to minimise the variance of the final map separately at different angular scales. The choice of using a blind component-separation method is motivated by an analogous study in \cite{Gainparametric}, which derive the requirements on the gain calibration for \textit{LiteBIRD} considering the parametric fitting \texttt{FGBuster}\footnote{\url{https://github.com/fgbuster/fgbuster}} pipeline. This latter analysis finds stringent requirements on the gain, especially for synchrotron- and dust-dominated frequencies.
Indeed, gain calibration uncertainties induce distortions in the foreground SEDs, which, if not adequately captured and described by the parametric modelling, bias the reconstructed CMB signal. Therefore, the parametric methods require a specific implementation to be able to marginalise over instrumental systematic effects. As an example, the \texttt{Commander} \cite{Commandergain} pipeline, (largely adopted for the analysis of \textit{Planck} data) jointly fits the foreground and instrumental parameters, thus being able to mitigate the impact of gain uncertainties on component-separation products. Since, as previously mentioned, the constraints on the gain calibration obtained in \cite{Gainparametric} appeared to be tight, we consequently aim to reproduce the procedure presented in \cite{Gainparametric} to set requirements on the gain calibration using the NILC minimum-variance component separation.  Specifically, we are interested in assessing to what extent the choice of a specific component-separation method affects the estimation of the tensor-to-scalar ratio in the presence of gain calibration uncertainties.

The paper is structured as follows. In \cref{Sim}, we outline the procedure to generate sky maps for each frequency of the \textit{LiteBIRD} satellite and to simulate the effect of an imperfect gain calibration.  \Cref{compsep} describes the NILC foreground cleaning method, its importance for studies of systematic effects and the specific algorithmic choices made for this analysis.
The procedure to estimate the tensor-to-scalar ratio from cleaned CMB maps and to set requirements on the gain calibration are explained in \cref{dr,require}, respectively. The results of this analysis are then presented in \cref{require}. In \cref{dis}, we summarise our main results and comment on future work.
\section{Simulation pipeline} \label{Sim}

In this section we describe our simulation framework. This includes the modelling of the sky emission, the generation of realistic observations by the instrument and the injection of relative polarisation gain uncertainty in the simulated maps. 

\subsection{Instrument model}
Contamination by Galactic emission demands that we consider observations over a broad frequency range. Therefore, \textit{LiteBIRD} is composed of three instruments: the Low Frequency Telescope \cite{LFT} (LFT); the Medium Frequency Telescope (MFT); and the High Frequency Telescope (HFT) \cite{MHFT}. The LFT is designed to observe CMB and synchrotron emission between 34 and 161\,GHz over 12 frequency channels. The MFT and HFT instruments (the so-called MHFT) will observe in the frequency ranges $89$-$224$ and $166$--$448$\,GHz, respectively. Such frequency coverage is designed to characterise the dust emission and increase the sensitivity in the CMB channels, corresponding to the frequency range $90 \lesssim \nu \lesssim 140$\,GHz. \textit{LiteBIRD} will operate with angular resolution ranging between $24$ and $71$ arc-minutes in order to cover the multipole range $2 \lesssim\ell \lesssim 300$ of the CMB $B$-mode angular power spectrum. The instrumental specifications used in this analysis are reported in \cref{Tablespec}.

\begin{table}
    \center
    \begin{tabular}{cccccc}
      \hline\hline
      Instrument & $\nu$ & Channel label & Beam FWHM &  Sensitivity & $N_{\text{bol}}$ \\
      & [GHz] & & [arcmin] & [\unit{\micro\kelvin}-arcmin] & \\
      \hline
      &40 & LFT-40 &70.5 & 37.42 & 48\\
      & 50 & LFT-50 &58.5 & 33.46 & 24 \\
      & 60 & LFT-60 &51.1 & 21.31 & 48 \\
      & 68 & LFT-68a &41.6 & 19.91 & 144 \\
      & 68 & LFT-68b &47.1 & 31.77 & 24\\
      & 78 & LFT-78a &36.9 & 15.55 & 144\\
      LFT &78 & LFT-78b & 43.8 & 19.13 & 48 \\
      & 89 & LFT-89a& 33.0 & 12.28 & 144 \\
      & 89 & LFT-89b & 41.5& 28.77 & 24 \\
      & 100& LFT-100 & 30.2 & 10.34 & 144 \\
      & 119& LFT-119 & 26.3 & 7.69 & 144 \\
      & 140& LFT-140 & 23.7 & 7.25 & 144 \\
      \hline
      & 100 & MFT-100 &37.8 & 8.48 & 366 \\
      & 119 & MFT-119&33.6 & 5.70 & 488 \\
      MFT & 140& MFT-140& 30.8 & 6.38 & 366\\
      & 166 & MFT-166&28.9 & 5.57 & 488 \\
      & 195 & MFT-195 &28.0 & 7.05 & 366 \\
      \hline
      & 195 & HFT-195 &28.6 & 10.50 & 254 \\
      & 235 & HFT-235 &24.7 & 10.79 & 254 \\
      HFT & 280 & HFT-280 & 22.5 & 13.80 & 254 \\
      & 337 &HFT-337& 20.9 & 21.95 & 254 \\
      & 402 &HFT-402& 17.9 & 47.45 & 338 \\
    \hline
    \end{tabular}
    \caption{\textit{LiteBIRD} instrumental specifications. From left to right: the instrument; the frequency channel and its label; the beam full width at half maximum (FWHM); the polarisation sensitivity; and the number of bolometers.}
    \label{Tablespec}
\end{table}

The first optical element of each telescope is a rotating half-wave plate (HWP) as polarisation modulator. The HWP's purpose is to reduce the contribution of $1/f$ noise and mitigate some other systematic effects such as gain drifts or intensity-to-polarisation leakage \cite{PTEP}. The full description of the real behaviour of the HWP is subject to uncertainties that contribute to the total systematic budget \cite{Giardiello2022,Monelli2023,2023arXiv230800967P}. However, since we aim to assess the impact of gain calibration uncertainties only, we assume an ideal polarisation modulator that does not generate additional systematic artefacts.

\subsection{Sky model} \label{Skymodel}
The total sky emission is given by the superposition of the CMB and Galactic signals. As mentioned in \cref{int}, we consider only polarised dust and synchrotron emission. Polarised dust emission is modelled with a modified blackbody SED \cite{dustplanck}: 

\begin{equation}
{[I,Q,U]}_{\mathrm{dust}} = {A_{[I,Q,U]}}_{\mathrm{dust}} \left({\frac{\nu}{\nu_{\mathrm{ref}}}}\right)^{{\beta}_\mathrm{d}}  \frac{B(\nu,{T}_{\mathrm{d}})}{B(\nu_{\mathrm{ref}},{T}_{\mathrm{d}})} \,,
\end{equation}
with $\beta_\mathrm{d}$ the dust spectral index, $B(\nu,T)$  the blackbody spectrum, $T_\mathrm{d}$ the dust temperature, and $\nu$ the frequency. The quantity $\nu_{\mathrm{ref}}$ corresponds to the pivot frequency, which allows us to define a reference template $A_{[I,Q,U]}$ for polarised dust.

Synchrotron emission can be modelled with a power-law spectrum \cite{foreg}: 

\begin{equation}
{[I,Q,U]}_{\mathrm{syn}} = {A_{[I,Q,U]}}_{\mathrm{syn}}  \left({\frac{\nu}{\nu_{\mathrm{ref}}}}\right)^{{\beta}_\mathrm{s}} \,,
\end{equation}
with $\beta_\mathrm{s}$ the spectral index of synchrotron, $\nu_{\mathrm{ref}}$ a pivot frequency, and ${A_{[I,Q,U]}}_{\mathrm{syn}}$ the synchrotron emission template at frequency $\nu_{\mathrm{ref}}$.

In this analysis, we simulate synchrotron and dust maps at each \textit{LiteBIRD} frequency assuming the \texttt{s0} and \texttt{d0} emission models, as implemented in the \texttt{PySM}\footnote{\url{https://pysm3.readthedocs.io/}} package. In these models, both polarised dust and synchrotron spectral parameters are uniform across the sky and equal to ${\beta}_\mathrm{s}=-3,\ {\beta}_\mathrm{d}=1.54$, and $T_\mathrm{d}=20\,\text{K}$. Such values correspond to a sky-average of the spectral parameters as fitted to \textit{WMAP} and \textit{Planck} data \cite{synchspec,dustplanck}.  Although it is well-known that foreground spectral parameters vary across the sky \cite{dustplanck,spec,SPASS}, we adopt this simplified sky model to: (i) disentangle the impact of the complexity of the sky model from that of the systematic effect under study and (ii)  match the foreground model used in \cite{Gainparametric}, to compare the impact of {gain calibration uncertainties} on different component-separation approaches.

The CMB component is generated from the \textit{Planck} best-fit angular power spectrum \cite{PlanckPS} using the Code for Anisotropies of Microwave Background \texttt{CAMB} \footnote{\url{https://camb.readthedocs.io/}} \cite{CAMBpaper} with the following set of cosmological parameters:  $r = 0$, $A_{\mathrm{s}} = 2 \times {10}^{-9}$, $n_{\mathrm{s}} = 0.965$ and $\tau = 0.06$ with $A_{\mathrm{s}}$ and $n_{\mathrm{s}}$, respectively, being the amplitude and spectral index of the power spectrum of primordial scalar fluctuations and the parameter $\tau$ being the reionisation optical depth. In addition to CMB and Galactic emission, we generate realisations of instrumental noise. This latter is assumed to be white, isotropic and uncorrelated between frequency channels. We recall that the contribution of 1/$f$ noise is negligible, since we assume an ideal HWP that fully mitigates this effect. We produce noise realisations for each frequency band using the polarisation sensitivity values reported in \cref{Tablespec}. For each \textit{LiteBIRD} channel, the simulated CMB, synchrotron and dust maps are smoothed with the corresponding Gaussian FWHM shown in \cref{Tablespec} and then coadded together with a noise realisation. We finally bring all frequency maps to \textit{LiteBIRD}'s lowest angular resolution: \ang{;70.5;}.

Following this procedure, we thus obtain $22$ polarisation Stokes $Q$ and $U$ maps with a common angular resolution of \ang{;70.5;}. We do not integrate over the bandwidth, thus having monochromatic maps. {The choice of not adopting realistic bandpasses is motivated by the fact that it does not have any relevant impact on the employed blind component-separation approach.} The NILC algorithm demands input maps to be scalar, therefore we convert the obtained polarisation full-sky $Q$ and $U$ maps into $E$- and $B$-mode maps. A detailed description of this transformation is presented in \cite{Zaldarriaga_1997}. We focus our analysis on $B$-mode maps, since we aim to assess the impact of the polarisation gain mis-calibration on the measurement of the tensor-to-scalar ratio.

\subsection{Simulating the relative polarisation gain mis-calibration}\label{gainsys}
To introduce a {gain calibration uncertainty}, we adopt a simple framework in which each frequency map is multiplied by a frequency-dependent gain calibration factor $g_{\nu}$ assumed to be homogeneous and constant over time. The total signal at a given frequency, $d_\nu^{B}$, can be thus expressed in the following way: 
\begingroup
\large
\begin{equation}
d_\nu^B = g_\nu ({m^{B}_{\mathrm{\nu,cmb}}} + {m^{B}_{\mathrm{\nu,fg}}}  + n_\nu) \,,
\label{eq:miscal}
\end{equation}
\endgroup
where $m^{B}_{\mathrm{\nu, cmb}}$ and $m^{B}_{\mathrm{\nu,fg}}$ represent the CMB and foreground $B$-mode maps, respectively, while $n_\nu$ is the white noise.
In the case of a perfect gain calibration, $g_{\nu}=1$ for each frequency $\nu$. To simulate the relative calibration error, we generate, for each channel, a random value of $g_{\nu}$ from a Gaussian distribution with standard deviation $\Delta g_{\nu}$:
\begingroup
\large
\begin{equation} \label{pert}
 g_{\nu} = 1 + \mathcal{N}(0,\Delta  g_{\nu}).
\end{equation}
\endgroup
Throughout this paper, the $g_{\nu}$ calibration factors are sampled from the Gaussian distribution shown in \cref{pert} and the gain uncertainty $\Delta g_{\nu}$, setting the width of this distribution, is the parameter on which we aim to derive a requirement. 

\section{Component-separation algorithm} \label{compsep}
In this work, as anticipated in \cref{int}, we adopt the NILC pipeline to recover CMB $B$ modes from \textit{LiteBIRD} multifrequency simulated data. Such a method has already been largely employed in CMB data analysis, e.g.\ for \textit{WMAP} \cite{NILC} and \textit{Planck} \cite{compsepplanck}, and it will also be one of the foreground cleaning pipelines for other next-generation CMB experiments, such as Simons Observatory \cite{SO}. 

NILC falls in the category of blind component-separation methods, since it performs a reconstruction of the CMB signal without any assumptions on the foreground spectral properties. It thus represents a valuable alternative to parametric approaches, as it is not affected by spectral mismodelling  of the Galactic polarised emission, which may significantly bias the final estimate of the tensor-to-scalar ratio. Moreover, previous studies \citep{ILCcalib} have shown that ILC techniques, such as NILC, are only mildly affected by calibration errors in the low signal-to-noise regime, as will be the case for CMB $B$-mode reconstruction with all upcoming experiments. Therefore, NILC is expected to represent the optimal framework to build a reliable reconstruction of CMB $B$ modes with less restrictive requirements for \textit{LiteBIRD} on {gain calibration uncertainties} as appeared to be the case for those obtained in \cite{Gainparametric}. We thus apply NILC to the mis-calibrated maps (described in \cref{gainsys}) and set requirements on the $\Delta g_\nu$ parameter for each frequency channel $\nu$, by checking its impact on the retrieved CMB $B$-mode solution and the tensor-to-scalar ratio estimation. 

\subsection{Mathematical formalism}
\label{nilc_math}
The ILC approach  consists of reconstructing a cleaned CMB signal by linearly combining input data at different frequencies with frequency-dependent weights:
\begingroup
\large
\begin{equation} \label{ILC}
\tilde{X}_{\textrm{CMB}} = \sum_{\nu=1}^{N_\nu} {\omega}_{\nu}   X^{\nu}= \sum_{\nu=1}^{N_\nu} {\omega}_{\nu}   (a^{\nu}_{\textrm{CMB}}X_{\textrm{CMB}}+X^{\nu}_{\textrm{fg}}+X^{\nu}_{\textrm{n}})\,,
\end{equation}
\endgroup
with $X^{\nu}$ representing the input observations at frequency $\nu$ in a specific domain, $a^{\nu}_{\textrm{CMB}}$ the CMB SED and $a^{\nu}_{\textrm{CMB}}X_{\textrm{CMB}},\ X^{\nu}_{\textrm{fg}}$ and $X^{\nu}_{\textrm{n}}$ the corresponding CMB, foregrounds and instrumental noise contributions, respectively. The optimal weights ${\omega}_{\nu}$ are estimated with the constraint of preserving the CMB blackbody signal:
\begingroup
\large
\begin{equation}
\sum_{\nu=1}^{N_\nu} {\omega}_{\nu}  a^{\nu}_{\textrm{CMB}}X_{\textrm{CMB}} = X_{\textrm{CMB}} \,,
\end{equation}
\endgroup
and to minimise the output variance $\text{Var}(\tilde{X}_{\textrm{CMB}})$.

In pixel-based and harmonic ILC, $X^{\nu}$ corresponds, respectively, to input $B$-mode maps and harmonic coefficients \citep{ILC,HILC}. Therefore, in these cases, the variance minimisation accounts for either the full range of angular scales or all directions on the sky. However, the relative contribution of Galactic emission and instrumental noise is expected to vary across the sky and among multipole moments. It follows that the subtraction of contaminants in the recovered CMB $B$ modes can be augmented through a variance minimisation performed locally in both domains. This is implemented in NILC, which extends the ILC approach to the needlet domain \citep{NILC_dela,NILC}. Needlets are a specific wavelet system that guarantees simultaneous localisation of the deprojected field in both real and harmonic space. In practice, needlet deprojection of a $B$-mode map at frequency $\nu$, $d_\nu^B$, returns a set of needlet maps, $\beta^{\nu}_{j}$, obtained by filtering its harmonic coefficients, ${a}^{B}_{\ell m,\nu}$, with different harmonic window functions $b_{j}(\ell)$:
\begingroup
\large
\begin{equation}\label{Needletcoeff}
  \beta^{\nu}_{j}(\hat{\gamma})  = \sum_{\ell,m} \left[{a}_{\ell m,\nu}  b_{j}(\ell)\right]  Y_{\ell m}(\hat{\gamma}) \,,
\end{equation}
\endgroup
with $\hat{\gamma}$ a specific direction on the sky and $j$ the needlet scale. For each $j$, we sample a specific range of multipole moments, with lower values of $j$ corresponding to ranges of larger angular scale. The total number of needlet scales $N_j$ depends on the targeted resolution for the reconstruction of the CMB signal. The procedure outlined in \cref{Needletcoeff} is equivalent to a convolution of $d_\nu^B$ with $N_j$ different kernels associated with the needlet filters $b_{j}(\ell)$. Since such filters are band-limited, each needlet coefficient $\beta^{\nu}_{j}(\hat{\gamma})$ is only sourced by modes of $d_\nu^B$ in a specific range of angular scales and within a finite spatial domain around $\hat{\gamma}$. Therefore, in NILC the input needlet maps are linearly combined separately for each needlet scale $j$:

\begingroup
\large
\begin{equation} \label{lincomb}
\beta^{\text{NILC}}_{j}(\hat{\gamma}) = \sum_{\nu=1}^{N_\nu} {\omega}_{\nu}^{j}(\hat{\gamma})   \beta^{\nu}_{j} (\hat{\gamma})=\sum_{\ell,m} {a}_{\ell m,j}^{\text{NILC}} Y_{\ell m}(\hat{\gamma})\,,
\end{equation}
\endgroup
so as to locally minimise the variance of $\beta^{\text{NILC}}_{j}(\hat{\gamma})$, thus effectively reducing the contamination locally in both pixel and multipole domains. The NILC weights, ${\omega}_{\nu}^{j}(\hat{\gamma})$, can be easily computed from the input needlet covariance $C_{\nu \mu}^{(j)} (\hat{\gamma})=\langle \beta^{\nu}_{j}(\hat{\gamma})  \beta^{\mu}_{j}(\hat{\gamma}) \rangle$: 
\begingroup
\large
\begin{equation}\label{weightscov}
    {\omega}_{\nu}^{j}(\hat{\gamma})  = \frac{a^{\mu}_{\textrm{CMB}}{\left(C_{\nu \mu}^{(j)-1}(\hat{\gamma})\right)}}{a^{\mu}_{\textrm{CMB}}{\left(C_{\nu \mu}^{(j)-1}(\hat{\gamma})\right)}a^{\nu}_{\textrm{CMB}}}.
\end{equation}
\endgroup
We can immediately observe that the estimation of NILC weights does not require any modelling or a priori knowledge of foreground spectral properties. In the baseline NILC analysis, the sample average in the covariance computation is performed within Gaussian axisymmetric domains (whose size varies with the considered needlet scale) centred around each sky direction. This is the approach adopted in this work. If the Galactic emission turns out to be very complex, the choice of such domains can be optimised through taking into account a data-driven blind estimation of the spatial distribution of the spectral properties of Galactic $B$-mode foregrounds, as implemented in Multiclustering-NILC \citep{MCNILC}. 

Once the output variance is separately minimised at the different needlet scales through \cref{lincomb}, the final NILC CMB $B$-mode map is obtained by performing an inverse needlet transformation of the needlet solutions $\beta^{\text{NILC}}_{j}$. In practice, this is done by first convolving each $\beta^{\text{NILC}}_{j}$ by the corresponding kernel (associated with $b_{j}(\ell)$) and then summing all the obtained maps at the different needlet scales:
\begingroup
\large
\begin{equation}\label{cleanedcmb}
B_{\textrm{CMB}}^{\text{NILC}} = \sum_{j}\sum_{\ell,m}\left[ {a}_{\ell m,j}^{\text{NILC}} b_{j}(\ell)\right] Y_{\ell m}(\hat{\gamma}).
\end{equation}
\endgroup
By looking at the direct and inverse needlet transformations of  \cref{Needletcoeff,cleanedcmb}, it follows that $\sum_{j} b_{j}^2(\ell)=1$ for each multipole moment.

\begin{figure} 
    \centering
    \includegraphics[width=\textwidth]{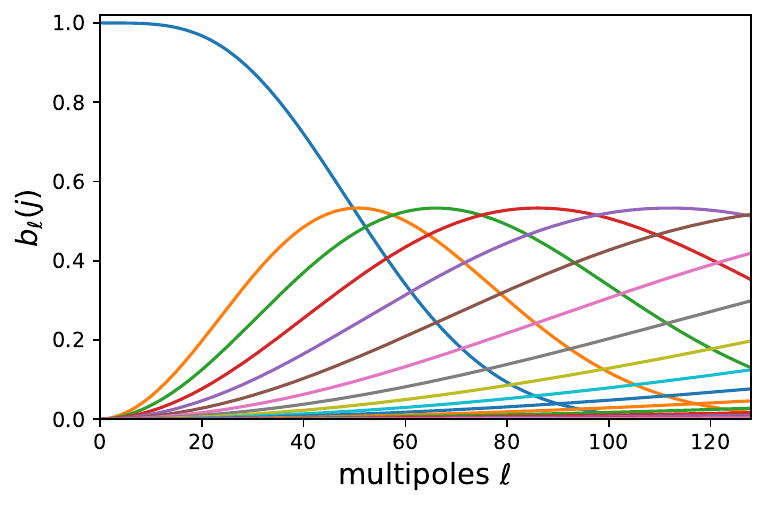}
    \caption{Configuration of Mexican-hat needlet bands adopted in this work. The needlet bands are shown in different colours for each needlet scale $j$. On the $x$-axis the multipoles $\ell$ with $\ell_\textrm{max} = 128$, while the $y$-axis, the needlet filters, $b_{j}(\ell)$, generated adopting a value of the needlet width parameter $B=1.3$. The first needlet band features a larger amplitude as it is obtained by merging together multiple bands according to \cref{eq:b_merge}.}
\label{needletband}
\end{figure}
\subsection{NILC implementation in our framework} \label{specNILC}
In this section, we provide details of the specific NILC configuration employed in this work. The main feature is the analytical form of the needlet functions $b_{j}(\ell)$. Commonly adopted constructions are the \emph{standard} \citep{doi:10.1137/040614359,2008MNRAS.383..539M}, the \emph{cosine} \citep{NILC} and the \emph{Mexican-hat} needlets \citep{2008arXiv0811.4440G}. In this analysis we adopt Mexican needlets generated with the Python module \texttt{MTNeedlet}\footnote{\url{https://javicarron.github.io/mtneedlet}}. {The Mexican needlet bands are Gaussian-shaped filters in harmonic space and their width is set by the parameter $B$}: lower values of $B$ correspond to a tighter localisation in harmonic space (fewer multipoles entering into any needlet coefficient), whereas larger values result in wider harmonic bands. We set $B=1.3$ to have adequate localisation in harmonic space at intermediate and small angular scales. At low multipoles, with such a choice of the value of $B$, needlet bands are so narrow that only a few modes would be sampled by each needlet scale. In this case, the estimation of the input needlet covariance in \cref{weightscov} would be highly uncertain and thus significantly deviate from a correct ensemble average. This induces a negative empirical correlation of noise and foreground residuals with the CMB signal in the output solution, known as the NILC \textit{bias} \citep{NILC_dela}. One of the main observable effects of NILC bias is a loss of power, quantified by ${C_{\ell}}^{\text{bias}}$, in the computed output power spectrum, $C_{\ell}^{\text{out}}$, with respect to the sum of contributions from single components:
\begingroup
\large
\begin{equation}\label{biasps}
     {C_{\ell}}^{\text{out}} -( {C_{\ell}}^{\text{CMB}} +  {C_{\ell}}^{\text{fg,res}} + {C_{\ell}}^{\text{n,res}})={C_{\ell}}^{\text{bias}}, 
\end{equation}
\endgroup
which is sourced by the negative cross-correlation terms. In  \cref{biasps}, ${C_{\ell}}^{\text{CMB}},  {C_{\ell}}^{\text{fg,res}}$ and ${C_{\ell}}^{\text{n,res}}$ represent angular power spectra of, respectively, the input CMB signal, foregrounds and noise residuals. If ${C_{\ell}}^{\text{CMB}}$ is much larger than $({C_{\ell}}^{\text{fg,res}} + {C_{\ell}}^{\text{n,res}})$, the angular power spectrum of the reconstructed CMB signal would be underestimated and this would impact the final inference of the cosmological parameters. The same effect also arises if the size of the spatial domain over which covariance is computed is too small. In order to overcome the NILC bias due to narrow needlet bands at low multipoles, we merge together the first $15$ needlet bands into a unique band as follows: 
\begingroup
\large
\begin{equation}
    b_0^{\text{new}}(\ell) = \sqrt{\sum_{j=0}^{14} {b}_{j}^2(\ell) } .
\label{eq:b_merge}
\end{equation}
\endgroup
The final configuration of harmonic needlet bands is shown in  \cref{needletband}. We adopt these filters to perform needlet deprojection of input multifrequency $B$-mode maps and then we linearly combine the needlet maps obtained to get the blackbody solution with lowest variance at each needlet scale.

\section{Procedure to set requirements} \label{dr}
In this section we summarise the procedure we use to find requirements on the relative polarisation gain calibration.
We describe how we compute the bias on the tensor-to-scalar ratio caused by the presence of this systematic effect and how we then set requirements.
\subsection{Tensor-to-scalar ratio bias} \label{drestimation}
To estimate the bias on $r$ due to gain mis-calibration, we apply the NILC algorithm on two distinct sets of maps, with the same CMB and noise realisations: one corresponds to the ideal calibration of the relative polarisation gain i.e.\ $\Delta g_\nu=0$ for all frequencies (see eq.~\ref{pert}); and the other uses a set of maps where the gain mismatch $(\Delta g_\nu \neq0)$ is injected in a specific frequency channel $\nu$. The component-separation step returns two different CMB solutions, noted as $m_\textrm{cmb}(\Delta g_\nu =0) $ and $m_\textrm{cmb}(\Delta g_\nu \neq0)$. We then apply the  \textit{Planck} \textit{GAL60} Galactic mask,\footnote{\url{https://pla.esac.esa.int}} which retains a sky fraction $f_{\text{sky}}=60\%$. 
For both sets, we compute the $B$-mode angular power spectrum of the residual maps, derived from the difference between output NILC solutions and input CMB map, thus obtaining 
\begingroup
\large
\begin{equation}\label{psres}
\begin{aligned} 
{C_{\ell}}^{\text{res}}(\Delta g_{\nu}=0)=     {C_{\ell}}^{\text{fg,res}}(\Delta g_{\nu}=0) + {C_{\ell}}^{\text{n}}(\Delta g_{\nu}=0)\,,\\[0.2cm]
{C_{\ell}}^{\text{res}}(\Delta g_{\nu}\neq 0)=     {C_{\ell}}^{\text{fg,res}}(\Delta g_{\nu}\neq 0) + {C_{\ell}}^{\text{n}}(\Delta g_{\nu}\neq 0) +{C_{\ell}}^{\text{cmb,dist}}\,,
\end{aligned}
\end{equation}
\endgroup
with $C_{\ell}^{\text{fg,res}}$ and $C_{\ell}^{\text{n}}$, respectively, the foregrounds and instrumental noise contribution after component separation, while ${C_{\ell}}^{\text{cmb,dist}}$ is a residual term associated with the distortion of the input CMB signal, which deviates from a perfect blackbody due to gain mis-calibration. This contribution is null in the case of an ideal gain calibration $\Delta g_\nu = 0$.
We call the $BB$ residual power spectra for the cases with and without mis-calibration, respectively,  $C_{\ell}^{\text{res}}(\Delta g_\nu \neq0)$ and $C_{\ell}^{\text{res}}(\Delta g_\nu =0)$. The angular power spectra are computed with the \texttt{anafast} routine implemented in the \texttt{healpy}\footnote{\url{https://healpy.readthedocs.io/}} module \cite{HEALpix}. This procedure does not account for correlations among multipoles induced by masking. However, as discussed in Ref.~\cite{PTEP}, this effect proves to have a negligible impact on the assessment of foreground and noise residual power spectra for large sky fractions, as considered in this paper. Moreover, in this case, we do not have to take $E$--$B$ leakage into account, since power spectra are computed directly on $B$-mode maps. 
To finally assess the impact of gain mis-calibration on the recovered CMB $B$ modes, we infer the tensor-to-scalar ratio $r$ from the output observed angular power spectrum ${C_{\ell}}^{\text{obs}}$, by adopting the following log-likelihood function \cite{likelihood,primB}:

\begingroup
\large
\begin{equation}\label{Likelihood}
-\text{ln}\mathcal{L}({C_{\ell}}^{\text{obs}}|r) = \sum_{\ell}\frac{2\ell+1}{2} {f}_\textrm{sky} \Bigg[\frac{{C_{\ell}}^{\text{obs}}}{{C_{\ell}}^{\text{th}}(r)} + \text{ln}({C_{\ell}}^{\text{th}}(r)) - \frac{2\ell-1}{2\ell+1}\text{ln}({C_{\ell}}^{\text{obs}})\Bigg]\,,
\end{equation}
\endgroup
where

\begingroup
\large
\begin{equation}\label{psobs}{C_{\ell}}^{\text{obs}} = {C_{\ell}}^{\text{res}}+ {C_{\ell}}^{\text{lensing}}, 
\end{equation} 
\endgroup
and the theoretical $BB$ power spectrum ${C_{\ell}}^{\text{th}}$ is given by

\begingroup
\large
\begin{equation}
{C_{\ell}}^{\text{th}}(r) = r{C_{\ell}}^{\text{GW},r=1} +{C_{\ell}}^{\text{lensing}} + {C_{\ell}}^{\text{res,eff}}.
\label{eq:cl_theo}
\end{equation}
\endgroup
In the equations above, $C_{\ell}^{\text{lensing}}$ is the $B$-mode spectrum induced by gravitational lensing, $C_{\ell}^{\text{GW},r=1}$ is the theoretical $B$-mode power spectrum sourced by tensor perturbations only for $r=1$, while the term 
$C_{\ell}^{\text{res,eff}}$
corresponds to a template model of noise and foreground residuals after component separation. Such a template power spectrum is obtained by applying NILC on $100$ simulations (with different CMB and noise realisations) of \textit{LiteBIRD} $B$-mode data in the case of ideal gain calibration and averaging over the corresponding $C_{\ell}^{\text{res}}(\Delta g=0)$.

We evaluate the likelihood function in \cref{Likelihood} over a grid of values of $r$ in the range $r\in [-1\times {10}^{-4},0.003]$ in steps of $\Delta r = 2\times{10}^{-7}$. We consider also $r<0$ values in order to allow for negative biases on $r$ (we recall that in all our simulations the CMB component is generated assuming $r=0$). Note that negative values of $r$ do not cause ${C_{\ell}}^{\text{th}}(r)$ to become negative in the logarithm of \cref{Likelihood} thanks to the presence of ${C_{\ell}}^{\text{lensing}}$ and ${C_{\ell}}^{\text{res,eff}}$ terms, which make the theoretical power spectrum ${C_{\ell}}^{\text{th}}(r)$ always positively defined.  We obtain the best-fit values of $r$ as the peak of the likelihood defined in \cref{Likelihood}, for both the mis-calibrated and ideal cases, and assess the bias due to the presence of this systematic effect, $\delta_r$, as
\begingroup
\large
\begin{equation}\label{bias}
    \delta_r = r(\Delta g_\nu \neq 0) - r(\Delta g_\nu = 0).
\end{equation}
\endgroup
We note that the inferred value of $r$ is driven by any deviation of an observed residual power spectrum with respect to a model ${C_{\ell}}^{\text{th}}$ (see eq.~\ref{eq:cl_theo}) in which no systematic effect is present. Therefore, the recovered tensor-to-scalar ratio in the ideal calibration case $r(\Delta g_\nu =0)$ is compatible with zero for each simulation.

\subsection{Summary of the procedure}
\label{sumproc}
We can now summarise the complete procedure employed to set requirements on the $\Delta g_\nu$ parameter for \textit{LiteBIRD}.
 
\begin{enumerate}

\item We generate the input multi-frequency maps by co-adding the simulated CMB signal, foregrounds and instrumental noise for each \textit{LiteBIRD} frequency channel, as described in \cref{Skymodel}.
\item Using \cref{pert}, we simulate the effect of an imperfect gain calibration at frequency $\nu$  by applying a gain calibration factor $g_\nu$ drawn from a Gaussian distribution whose standard deviation is given by the $\Delta g_\nu$ parameter. All other frequency maps are therefore left unaltered. At this stage, for each realisation of CMB and noise components, we have two sets of maps: an ideal one with perfect gain calibration $m(\Delta g_\nu =0)$; and one that includes the systematic effect $m(\Delta g_\nu \neq0)$

\item We apply the NILC component-separation procedure to both simulated data sets. This step returns two distinct CMB $B$-mode solutions in the pixel domain, $m_{cmb}(\Delta g_\nu =0)$ and  $m_{cmb}(\Delta g_\nu \neq0)$; we then apply a 60$\%$ sky cut to mask the Galactic plane.

\item  We compute the residual power spectra ${C_{\ell}}^{\text{res}}(\Delta g_\nu=0)$  and ${C_{\ell}}^{\text{res}}(\Delta g_\nu \neq0)$, as defined in \cref{psres}. Using the likelihood function in  \cref{Likelihood}, we derive the best-fit tensor-to-scalar ratio in both cases, $r(\Delta g_\nu=0)$ and $r(\Delta g_\nu \neq 0)$, and compute the bias $\delta_r$ as in \cref{bias}.
    
\item We repeat the four steps above for different values of $\Delta g_\nu $. For all $\Delta g_\nu$ values, we consider $100$ simulations with different realisations of CMB, noise and $g_{\nu}$, obtaining therefore $100$ values of $\delta_r$.
    
\item We build the histogram of the bias distribution for each value of  $\Delta g_\nu$ and estimate the quantity $\Delta = \sqrt{{\mu}^{2} + {\sigma}^{2}}$ of that distribution (see \cref{drdg}) with $\mu$ the mean value and $\sigma$ the standard deviation. {The $\Delta$ quantity is equivalent to the root mean square (RMS) of the distribution}.

\item  We derive an empirical relation for $\Delta $ as a function of $\Delta g_\nu$.

\item From this relation, we obtain the requirement on the gain calibration, ${\Delta g}_{\textrm{req}}$, for each frequency $\nu$. This corresponds to the value of $\Delta g_\nu$ for which $\Delta$ is equal to the gain systematic budget allocated to a single channel $\delta_r^{\textrm{req}}$, $\delta_r^{\textrm{req}}= 6.5 \times {10}^{-6}/22$ \footnote{The total budget assigned to the gain systematic effects for \textit{LiteBIRD}, equal to $\Delta r = {6.5} \times {10}^{-6}$, is uniformly distributed over all frequency channels. Therefore, the budget associated to a single frequency channel is equal to $\delta_r^{\textrm{req}}= 6.5 \times {10}^{-6}/22$. } \cite{PTEP}.
    
\item We repeat steps 1 to 8 for all \textit{LiteBIRD} frequency channels and obtain ${\Delta g}_{\textrm{req}}(\nu)$, which represent the requirements on each single frequency channel.
\end{enumerate}

\section{Results} 
\label{require}
In the following we summarise our results, including the requirements obtained on the single frequency channels (\cref{drdg}) and the impact on the tensor-to-scalar ratio when all frequency channels are mis-calibrated simultaneously (\cref{all_miscal}).
In \cref{drdg}, the systematic effect is propagated through a single frequency channel, considering different amplitudes of the gain calibration uncertainty $\Delta g_{\nu}$ (eq.~\ref{pert}). The requirement on $\Delta g_{\nu}$, being the value for which the single-channel gain systematics budget ($\delta_r^{\textrm{req}}= 6.5 \times {10}^{-6}/22$) is met, is then derived for each \textit{LiteBIRD} frequency. In \cref{all_miscal}, the systematic error is injected into all frequency channels simultaneously, their respective gain calibration factors being sampled from Gaussian distributions whose widths are set by the requirements per-channel ${\Delta g}_{\textrm{req}}(\nu)$ derived in \cref{drdg}. By doing so, we assess whether the total budget allocated to gain systematics ($\Delta r = 6.5 \times {10}^{-6}$) is fulfilled, considering the previously derived set of requirements. In this analysis, for simplicity, we assume the gain calibration errors to be uncorrelated among the frequency channels. In a more realistic situation these errors are supposed to be partially correlated due to the gain calibration procedure and this is expected to lead to less stringent constraints on the gain calibration. Therefore, the derived set of requirements is considered a conservative one.

\subsection{Single frequency requirements}\label{drdg}

\begin{figure}
    \centering
    \includegraphics[width=\textwidth]{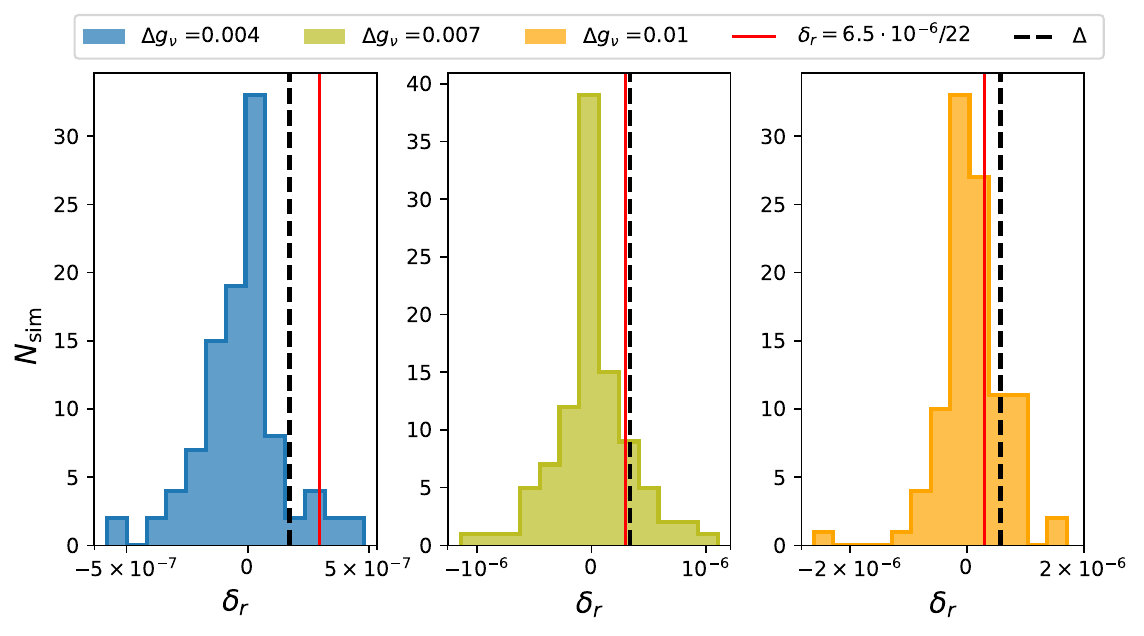}
    \caption { Distribution of $\delta_r$ (for $100$ different simulations) when the \textit{LiteBIRD} LFT $100$-GHz frequency maps are affected by gain calibration uncertainties of $\Delta g_\nu=0.004$ (left), $0.007$ (centre) and $0.01$ (right); see  \cref{eq:miscal,pert}). The red solid vertical line represents the gain systematic budget per channel ${\delta_r}^{req} = 6.5 \times {10}^{-6} /22$, while the dashed black vertical line shows the value of $\Delta = \sqrt{{\mu}^2 + {\sigma}^2}$.}
    \label{hist_plot}
\end{figure}

The definition of requirements on the gain calibration accuracy relies on the choice of a unique quantity derived from our $\delta_r$ distribution, which should encompass all the statistical variations due to a gain systematic effect. One could look for the value of $\Delta g$ for which the mean value of $\delta_r$ coincides with the assigned budget. However, we observe that for increasing values of $\Delta g$, the mean value of $\delta_r$ fluctuates around zero. Indeed, {gain calibration uncertainties} have a negligible impact on average, this being expected because the values of the gain calibration factor $g$ are drawn from a Gaussian distribution centred at one. On the other hand, we clearly note an effect of {gain calibration uncertainties} on a single simulation, which translates into a larger standard deviation of the distribution, as we strengthen the amplitude of the perturbation (see \cref{hist_plot}). We therefore take as a reference the quantity $\Delta = \sqrt{{\mu}^{2}+{\sigma}^{2}}$ derived from the $\delta_r$ distribution, this latter taking into account both the \textit{bias} $\mu$ (representing the deviation of the $\delta_r$ mean value from zero) and the \textit{extra variance} $\sigma$. Note that this extra variance is induced by the presence of gain calibration uncertainties only and does not contain 
contributions from the noise variance and foreground residual uncertainties due to component separation.  Indeed, as mentioned in \cref{sumproc}, the $\delta_r$ distribution is derived from a difference of $r$ in ideal calibration and mis-calibrated cases therefore the contribution to the variance from the noise and foreground residuals cancels out.
In our analysis, the contribution from the bias is negligible and we could equivalently use only the standard deviation of the distribution as a reference quantity. Such a statement is not necessarily true for other systematic effects or component-separation techniques, however, and therefore we use the $\Delta$ quantity to remain general.

\begin{figure}
    \centering
    \includegraphics[width=\textwidth]{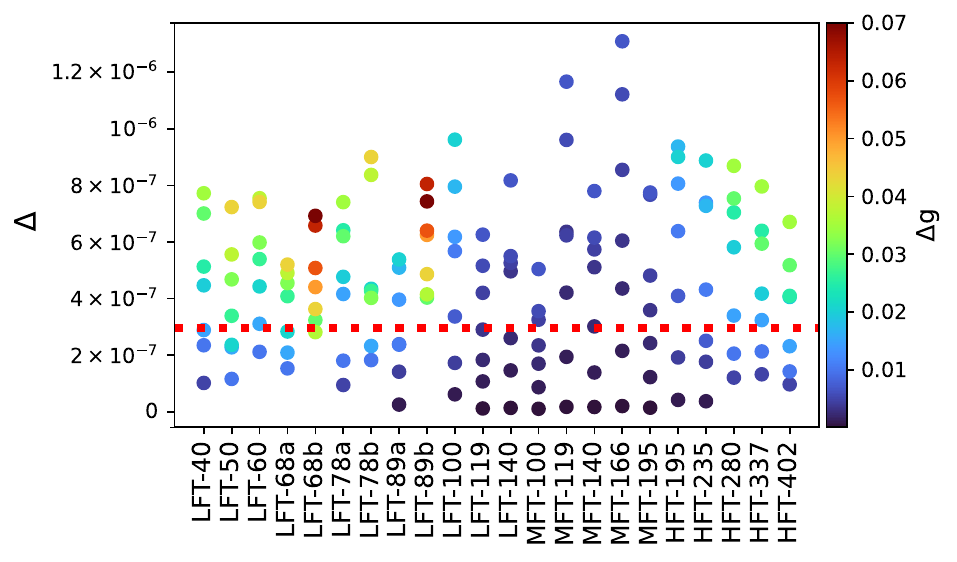}
    \caption{Evolution of $\Delta$  $(= \sqrt{{\mu}^{2} + {\sigma}^{2}}$) as a function of {the gain calibration uncertainty} $\Delta g_\nu$ for all \textit{LiteBIRD}'s channels. The red dashed horizontal  line corresponds to the gain systematic budget per channel ${\delta_r}^{\textrm{req}} = 6.5. {10}^{-6}/22$.}
    \label{drdgplot}
\end{figure}

We perturb independently each channel, considering several values of $\Delta g_\nu$ (varying in a common range for all channels) and translate it into $\Delta$ variations. From this, we can determine the sensitivity of each channel under gain calibration uncertainties, thus allowing us, for the rest of the analysis, to adapt and tighten the range of $\Delta g_\nu$ values (linearly spaced) for each frequency. Afterwards, we perform the steps $1$ to $6$ described in \cref{sumproc} and obtain for each \textit{LiteBIRD} frequency channel, a distribution of $\delta_r$ for eight different {gain calibration uncertainties} $\Delta g_\nu$, varying in different ranges. As an example, \cref{hist_plot} shows the bias distribution for the LFT $100$-GHz channel for three different values of $\Delta g_\nu$, namely 0.004, 0.007 and 0.01. From each distribution we derive the mean value of the bias, $\mu({\delta}_r)$ and the standard deviation $\sigma({\delta}_r)$ and compute $\Delta$, for each value of $\Delta g$. The amplitude of $\Delta $ as a function of the {gain calibration uncertainty} for each frequency channel is shown in \cref{drdgplot}. As expected, at all frequencies the amplitude of $\Delta$ increases with the {gain calibration uncertainty}. We observe that the largest values of $\Delta $ are obtained for the central frequency channels, which correspond to those having larger NILC weights. Indeed, for small values of $\Delta g$ ($\lesssim 4 \times {10}^{-2}$) the assigned budget is exceeded at CMB frequencies (from LFT-119 to MFT-195), while the low and high-frequency channels are less senstive and allow larger amplitudes of the gain perturbation.  This trend is opposite to that found in Ref.~\cite{Gainparametric} and highlights the different impact of gain mis-calibration when considering different component-separation approaches. 

\begin{figure}
    \centering
    \includegraphics[width = \textwidth]{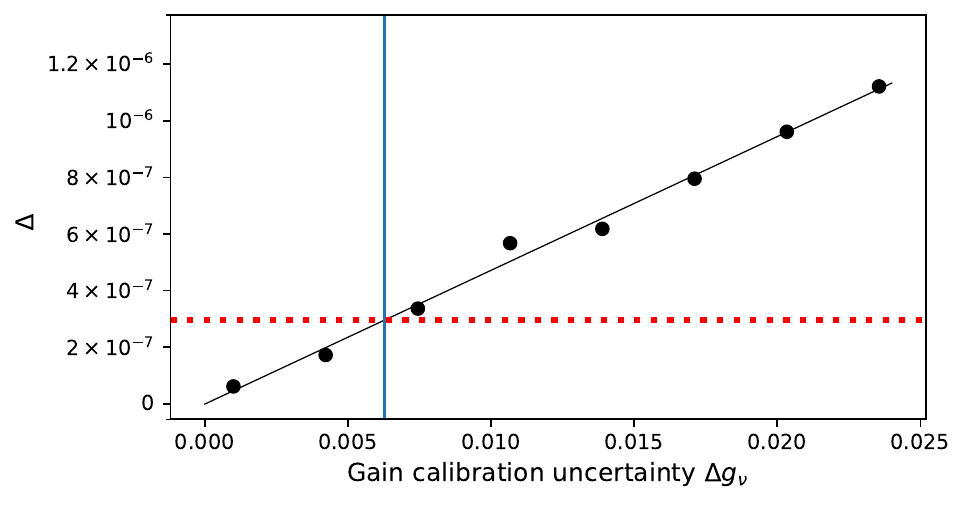}
    \caption{Variation of $\Delta$ ($= \sqrt{{\mu}^{2} + {\sigma}^{2}}$) with the gain calibration uncertainty $\Delta g_\nu$ for the LFT $100$-GHz channel. The red dashed line shows the single channel gain systematic budget ${\delta_r}^{\textrm{req}} =6.5 \times {10}^{-6} /22$ and the blue solid line corresponds to the requirement on the gain calibration accuracy obtained for the LFT 100-GHz channel, namely 0.63\%.}
    \label{req100}
\end{figure}

Once the $\Delta$ values as a function of $\Delta g_\nu$ are obtained for each channel, we perform an interpolation in order to obtain an empirical relation among these two quantities. 
This is well approximated by a linear function for all frequency channels. 
These interpolating functions are then employed to derive the requirement on the gain calibration for each \textit{LiteBIRD} frequency channel as the value of  ${\Delta g }_{\nu}$ for which $\Delta = \delta_r^{\textrm{req}} = 6.5 \times {10}^{-6} / 22$,  the budget allocated to a single frequency channel.  The latter is derived by uniformly distributing \textit{LiteBIRD}'s total gain systematic budget ($\Delta r = 6.5 \times {10}^{-6}$) over the $22$ distinct frequency channels.  As an example, we show in \cref{req100} the amplitude of $\Delta$ as a function of $\Delta g_\nu$ for the LFT 100-GHz channel, together with the corresponding interpolating function. The  intercept with the horizontal line $\delta_r^{\textrm{req}}$ allows us to determine ${\Delta g }_{\text{req}}$. 

We report the derived requirements per channel ${\Delta g}_{\textrm{req}}(\nu)$ in \cref{LITEBIRDPAR}.
We find that the requirements on $\Delta g_\nu$ range between $0.16$ and $ 3.22 \%$. Such values are less restrictive (by two orders of magnitude) than those obtained when the \texttt{FGBuster} parametric component-separation pipeline is applied to an analogous simulated \textit{LiteBIRD} data set, as found in Ref.~\cite{Gainparametric}.
We recall that NILC, and blind methods more generally, construct a linear combination of frequency maps in order to recover th CMB blackbody signal. It follows that the weights of the linear combination are larger at frequencies where the CMB is less obscured by other sources of emission, and they tend to be smaller at frequencies where foregrounds are more dominant. Therefore, {gain calibration uncertainties} have a larger impact in the frequency range where NILC weights are larger, thus leading to more stringent requirements around CMB frequencies. We then expect a correlation between NILC weights and the channels sensitivity to {gain calibration uncertainties}, quantified in terms of the requirement on $\Delta g_\nu$. 

\begin{table}
    \begin{center}
    \begin{tabular}{cc}
     \hline\hline
     \noalign{\vskip1pt}
      Channel Label&  Single channel ${\Delta g}_{\textrm{req}}(\nu) [\%]$ \\ [0.5ex] 
     \hline
     LFT-40 & 1.36   \\ 
     LFT-50 & 2.19    \\
     LFT-60 & 1.51     \\
     LFT-68a & 2.12  \\
     LFT-68b & 3.22   \\
     LFT-78a & 1.27  \\
     LFT-78b & 1.72    \\
     LFT-89a & 1.04  \\
     LFT-89b & 2.05  \\
     LFT-100 & 0.63  \\
     LFT-119 & 0.33 \\
     LFT-140 & 0.25 \\
     MFT-100 & 0.43  \\
     MFT-119 & 0.17 \\
     MFT-140 &  0.23  \\
     MFT-166 & 0.16  \\
     MFT-195 & 0.26  \\
     HFT-195 & 0.52   \\
     HFT-235 & 0.70  \\
     HFT-280 & 1.18  \\
     HFT-337  & 1.31  \\
     HFT-402 & 1.70  \\
     \hline
    \end{tabular}
    \end{center}
\caption{Requirements on the relative polarisation gain  calibration for all \textit{LiteBIRD} frequency channels. {This set of requirements is obtained assuming the \texttt{d0s0} sky model.}}

\label{LITEBIRDPAR}
\end{table}

\begin{figure}
    \centering
    \includegraphics[width =\textwidth]{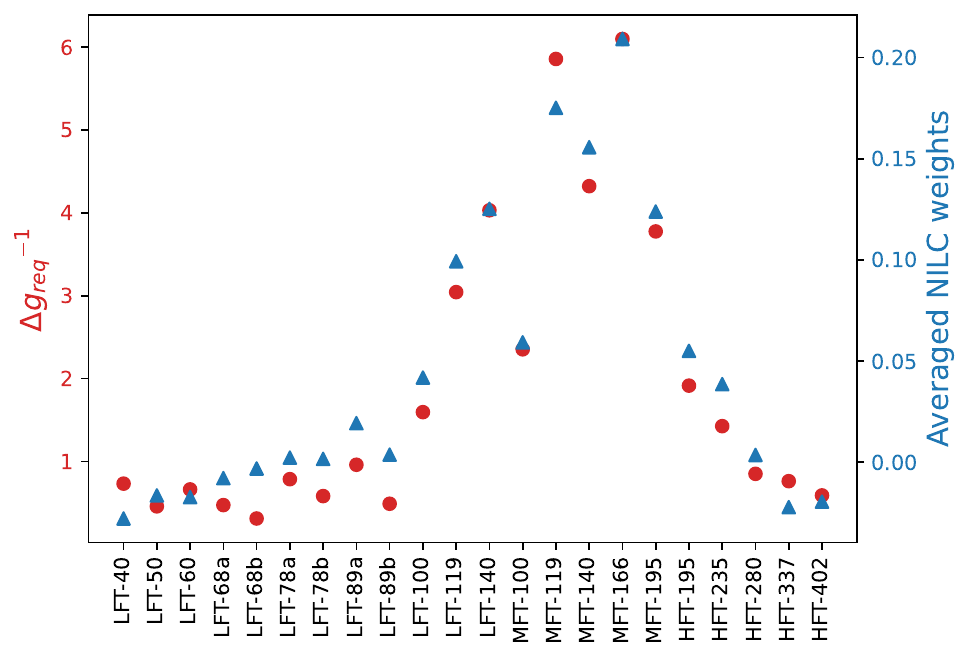}
    \caption{Trend of the inverse of the requirements on the relative polarisation gain calibration ${\Delta g}_{\textrm{req}}^{-1}$ (red points), reported in \cref{LITEBIRDPAR}, and averaged NILC weights in the first needlet band ${\bar{\omega}}_{\nu}^0$ (blue triangles) for the different \textit{LiteBIRD} frequency channels.}
    \label{corrweightsreq}
\end{figure}

In \cref{corrweightsreq}, we show the inverse of the requirements, ${\Delta g}_{\textrm{req}}^{-1}(\nu)$, and the average NILC weights in the first needlet band (${\bar{\omega}}_{\nu}^0$), as a function of the frequency. We can see that the frequency dependence of ${\Delta g}_{\textrm{req}}^{-1}(\nu)$ is strongly correlated with that of ${\bar{\omega}}_{\nu}^0$. We report the correlation with the NILC weights in the first needlet band $j=0$, since this band is the one that samples modes at the largest angular scales and has more constraining power on $r$. We observe correlations also for the other needlet scales. The trend observed in \cref{corrweightsreq} also explains that for the channels observing at the same frequency but with different sensitivities, we observe more stringent requirements for the \textit{a}-channels than for the \textit{b}-channels.  The \textit{b}-channels possessing a higher noise level, tend to be down-weighted (with respect to \textit{a}-channels) in the NILC process, and therefore are less sensitive to gain calibration uncertainties.

\subsection{Simultaneous mis-calibration of all channels}
\label{all_miscal}

After obtaining requirements separately for each frequency channel, we assess whether the impact on the tensor-to-scalar ratio is compatible with the total budget allocated to {gain calibration uncertainties} when all frequency channels are mis-calibrated assuming the ${\Delta g}_{\textrm{req}}(\nu)$ values reported in \cref{LITEBIRDPAR}. 
Each frequency channel is affected by a specific gain calibration factor derived from a Gaussian distribution whose standard deviation corresponds to the requirements ${\Delta g}_{\textrm{req}}(\nu)$ shown in \cref{LITEBIRDPAR}.

\begin{figure}
    \centering
    \includegraphics[width=\textwidth]{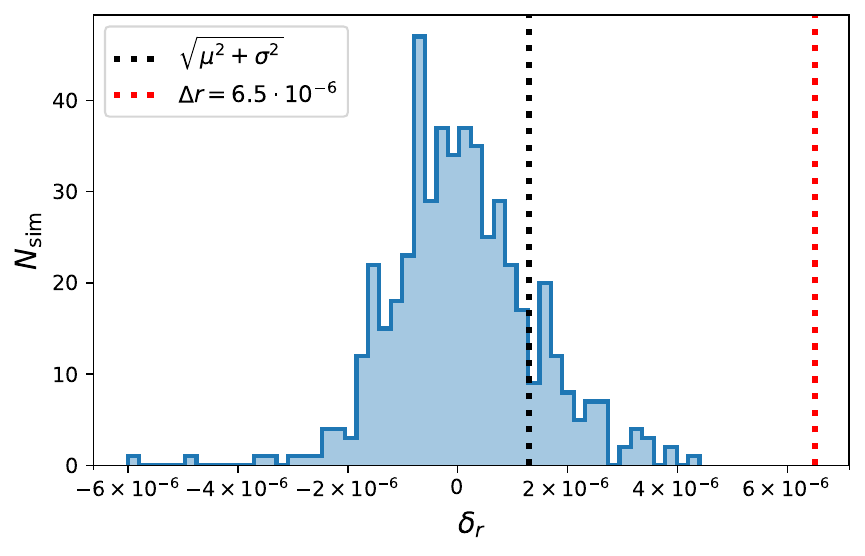}
    \caption{Distribution of $\delta_r$ for $N_{\textrm{sim}} = 500 $ when all channels are perturbed simultaneously with their corresponding requirement ${\Delta g}_{\textrm{req}}(\nu)$ (reported in \cref{LITEBIRDPAR}). The dotted red line indicates the budget allocated to gain systematics $\Delta r = 6.5 \times {10}^{-6}$ and the dotted black line represents the value of $\Delta =  \sqrt{{\mu}^{2}+{\sigma}^{2}}$ of the distribution.}
    \label{gaincomb}
\end{figure}

We run the NILC component-separation process on a set of $500$ map realisations, and, for each of them, we determine $\delta_r$ value as the difference between the tensor-to-scalar ratio $r$ for the perturbed and unperturbed case (see eq.~\ref{bias}). The distribution of $\delta_r$ among all 500 realisations is shown in \cref{gaincomb}. The resulting value of $\Delta \approx 1.29\times{10}^{-6}$ is lower than the \textit{LiteBIRD} gain systematics budget $\Delta r= 6.5 \times {10}^{-6}$ \cite{PTEP} by a factor of approximately $5$. This result shows that the biases on the tensor-to-scalar ratio $\delta_r$ originating from separately mis-calibrating all frequency channels by their corresponding ${\Delta g}_{\textrm{req}}(\nu)$,  do not add up linearly if the same mismatches are applied simultaneously to all channels.

\begin{figure}
    \centering
    \includegraphics[width = \textwidth]{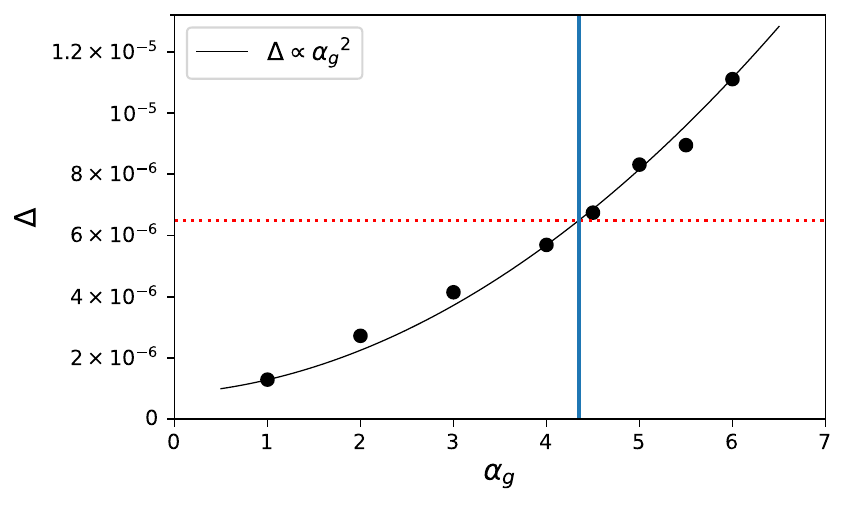}
    \caption{{The quantity $\Delta = \sqrt{{\mu}^{2} + {\sigma}^{2}}$ as a function of $\alpha_g$. The red dotted line shows the total gain systematics budget $\Delta r= 6.5 \times {10}^{-6}$, while the blue solid line corresponds to the value of $\alpha_g$ by which the requirements on \texttt{d0s0} should be rescaled to match the total gain systematics budget. }}
    \label{alphag_d0s0}
\end{figure}

This result originates from the adaptive behaviour of the NILC weights, which automatically tend to readjust themselves to the different frequency scalings of the sky components. Such re-adjustment of the weights induces correlations in the impact of mis-calibration of different channels, which does not allow a linear addition of the contributions from single channel mismatches.  \Cref{gaincomb} shows that a considerable margin is available before reaching
the threshold bias allocated to gain mis-calibration, suggesting that such requirements may be revisited, e.g.\ by relaxing the most stringent ones and reducing those for less sensitive channels. Another possibility is to apply a common multiplicative factor, $\alpha_g$, to the set of requirements presented in \cref{LITEBIRDPAR} and simultaneously mis-calibrate all channels by these rescaled ${\Delta g}^{\prime}_{\textrm{req}}(\nu)=\alpha_g {\Delta g}_{\textrm{req}}(\nu)$ values. To do this, we consider multiple values of $\alpha_g$ in the range $[1,6]$. We then interpolate the evolution of  $\Delta$ derived from the  ${\delta}_r$ distributions as a function of $\alpha_g$, and estimate the $\alpha_g$ value that leads to a total bias of $\Delta r= 6.5 \times {10}^{-6}$. \Cref{alphag_d0s0} indicates that the ${\Delta}$ dependence on $\alpha_g$ is quadratic. Such a trend is explained by the fact that, since we apply a common scaling factor to all channels simultaneously, the amplitude of output residuals is expected to scale linearly with $\alpha_g$ and therefore the output $BB$ power spectrum will be proportional to ${\alpha_g}^{2}$.  

{Using the interpolating function shown in \cref{alphag_d0s0}}, we find that the total gain systematics budget is reached for $\alpha_g \approx 4.4$. We are therefore able to set an upper limit on the relative polarisation gain requirements, corresponding to an amplification by this factor of the ${\Delta g}_{\textrm{req}}(\nu)$ values reported in \cref{LITEBIRDPAR}. Overall, the available margin between the obtained $\Delta$ and threshold values could possibly absorb larger impacts of {gain calibration uncertainties} in scenarios with more complex foregrounds, as shown in the next section.

\subsubsection{{Extension to more complex sky models}}\label{newskymodels}

We now turn to investigating the impact on $\delta_r$ if the requirements presented in \cref{LITEBIRDPAR} are applied simultaneously to frequency maps, but assuming more complex foreground models.

\paragraph{1. Sky model}
Recall that the results presented in \cref{all_miscal} are obtained considering a sky model with constant synchrotron and dust spectral indices across the sky (\texttt{d0s0} model) and using NILC component separation. The \texttt{s0} and \texttt{d0} models are, however, a simplification of the actual Galactic emission and constitute the lowest level of complexity in the foreground modelling provided by the \texttt{PySM} package.

The first option would be to set requirements on the gain calibration following the procedure presented in \cref{sumproc} for more complex sky models, these being as realistic as possible in order to mimic the conditions of real polarisation observations. This would, however, need an extremely accurate knowledge of the Galactic foreground spectral energy distribution. Such knowledge has not yet been achieved but investigating it is generating much interest in the CMB community \cite{Mangilli_2021,Vacher2022,Ritacco_2023}.

\begin{figure}
    \centering
    \includegraphics[width =\textwidth]{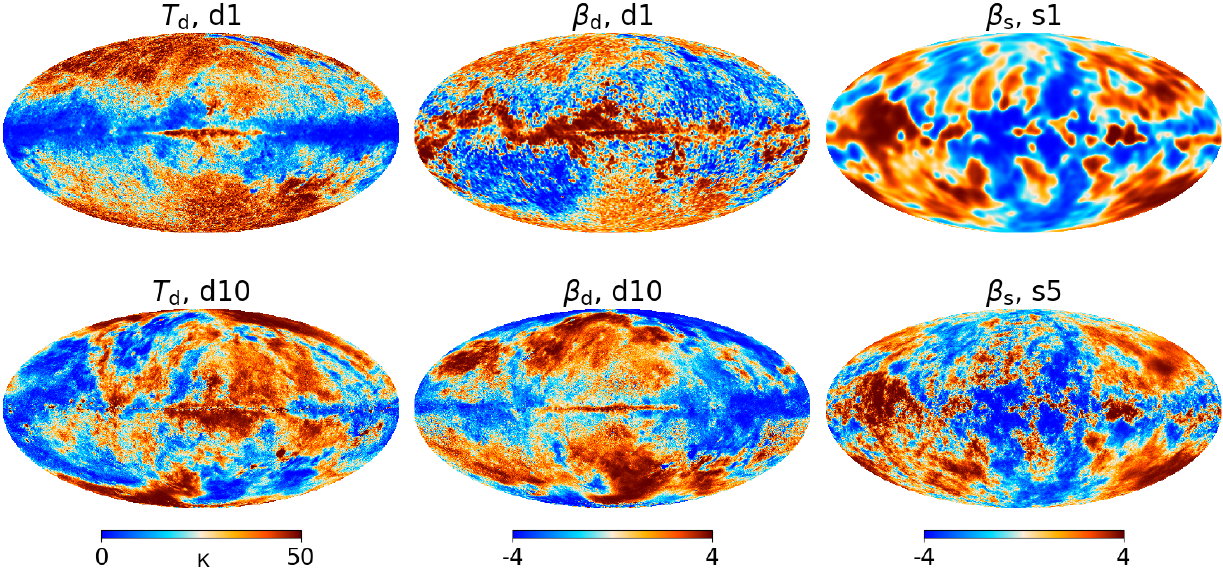}
    \caption{{Polarised dust and synchrotron spectral indices $\beta_\mathrm{d} $, $T_\mathrm{d}$ and $\beta_\mathrm{s}$ for  \texttt{d1s1} (top) and \texttt{d10s5} (bottom) sky models. These maps are generated for a resolution parameter of $N_\textrm{side}=512$. }}
    \label{spectral_indices}
\end{figure}

In this section, we aim to assess if the requirements derived for the \texttt{d0s0} foreground model are robust against more complex foreground scenarios. We thus consider two additional \texttt{PySM} sky models: \texttt{d1s1} and \texttt{d10s5}. In the \texttt{d1s1} model, the dust and synchrotron spectral indices vary across the sky. The dust template corresponds to the 353-GHz map from  \textit{Planck} \cite{Planck2015,dustplanck} and the dust spectral parameters maps are obtained by applying the \texttt{Commander} pipeline \cite{compsepplanck} to the  \textit{Planck} data set. The synchrotron template corresponds to the \textit{WMAP} 9-year 23-GHz $Q$/$U$ maps \cite{WMAP} and the spectral index map is obtained by combining the \textit{Haslam} 408-MHz data and \textit{WMAP} 23-GHz 7-year data \cite{synchspec}. In the \texttt{d10s5} model, the synchrotron spectral index map is rescaled to account for the larger variability observed by the S-PASS experiment \cite{SPASS}, which mapped synchrotron emission at $2.3$\,GHz. The maps of thermal dust spectral parameters are obtained by applying the GNILC component-separation technique to the \textit{Planck} data set. {The variations of polarised dust and synchrotron spectral indices $T_\mathrm{d}$, $\beta_\mathrm{d}$ and $\beta_\mathrm{s}$ across the sky are shown in \cref{spectral_indices} for both sky models described above.}

\paragraph{2. MC-NILC component separation}
In a framework with complex foreground emission, the NILC pipeline, which performs simple local variance minimisation across the sky, may be suboptimal, since it is not enginereed to fully handle the local spectral variations of foregrounds. We therefore use the MC-NILC (Multi-Clustering Needlet ILC) \cite{MCNILC} foreground-cleaning method which aims at minimising the variance within sky patches, also called \textit{clusters}. MC-NILC accounts for the spatial variability of spectral properties of foreground $B$ modes by identifying a tracer of their distribution across the sky, with a limited number of a priori assumptions.
The spectral distribution of dust and synchrotron spectral parameters are assessed by computing the ratio of foreground $B$-mode maps at two distinct frequencies: one at high frequency ($337$\,GHz) and one at CMB frequencies ($119$\,GHz).  Such a ratio allows us both to estimate an effective thermal dust spectral index and an emission ratio of synchrotron and dust at the CMB-dominated frequency.

In this analysis, we consider two distinct approaches of MC-NILC. Firstly, the \textit{ideal} approach where clusters are built for each needlet scale from the ratio of the input $337$-GHz and $119$-GHz foreground-only $B$ modes (noiseless). Note that the ideal MC-NILC approach is not data-driven because the foreground $B$-mode templates are directly derived from simulations. However, such an approach remains helpful to assess the maximal capability of MC-NILC to perform foreground subtraction, since the ratio of simulated foreground $B$ modes is able to trace, in an optimal way, the spatial variations of spectral indices across the sky. In the context of gain calibration, the ideal MC-NILC allows us to derive the impact of {gain calibration uncertainties} on more complex foreground modelling when the employed component-separation method is optimal.  Secondly, we consider  \textit{realistic} MC-NILC, where templates of foreground $B$ modes at the two frequencies of interest (for the construction of the tracer) are obtained by applying the Generalised Needlet ILC (GNILC) \cite{genILC} formalism to observed multi-frequency data. The patches are then built from a unique emission ratio at $337$ and $119\,$GHz of foreground $B$ modes, for all needlet scales. The realistic MC-NILC has the benefit of being able to build a tracer of the spectral variations of foregrounds directly from observed multi-frequency data. However, it suffers from residual contamination of CMB and noise in the GNILC foreground templates and therefore the reconstruction of the CMB signal in the end is not as optimal as in the ideal MCNILC approach.

In this study, we consider 50 clusters of equal area, the optimal number of clusters being assessed by comparing the foreground residuals and the bias in the CMB reconstruction after performing MC-NILC on the \textit{LiteBIRD} data set with different numbers of clusters. A detailed description of the sky-patch optimisation is presented in Ref.~\cite{MCNILC}. The variance minimisation is then performed within each cluster using the NILC component separation, considering the same configuration as the one presented in \cref{specNILC}.

\paragraph{3. Simulations and results}

We simulate multi-frequency maps of CMB, dust, synchrotron and noise, as described in \cref{Sim}, considering the two different foreground modelling schemes presented above. We propagate the gain calibration uncertainties to all channels simultaneously, according to the requirements in \cref{LITEBIRDPAR} and apply MC-NILC to a set of $500$ realisations. Since we aim to assess the impact of the gain mis-calibration on each simulation, we also apply MC-NILC to the twin set of maps without systematic effects.

We apply a common mask to the output CMB maps, corresponding to the previously used \texttt{GAL60}  \textit{Planck} mask with an additional 10$\%$ obtained by thresholding the averaged foreground residuals map (over 500 simulations, without gain mismatch) smoothed with a $\textrm{FWHM}=3^\circ$ Gaussian beam. An analogous masking strategy is employed in Ref.~\cite{PTEP}, and retains a $50\%$ fraction of the sky. Note that such a masking strategy cannot be employed to analyse real data because it requires a foreground residuals template not directly accessible from observations. However, as shown in Ref.~\cite{MCNILC}, very similar results are obtained with a fully data-driven approach where a template of foreground residuals is derived by combining the MC-NILC weights with the GNILC frequency maps.

\begin{figure}
    \centering
    \includegraphics[width =\textwidth]{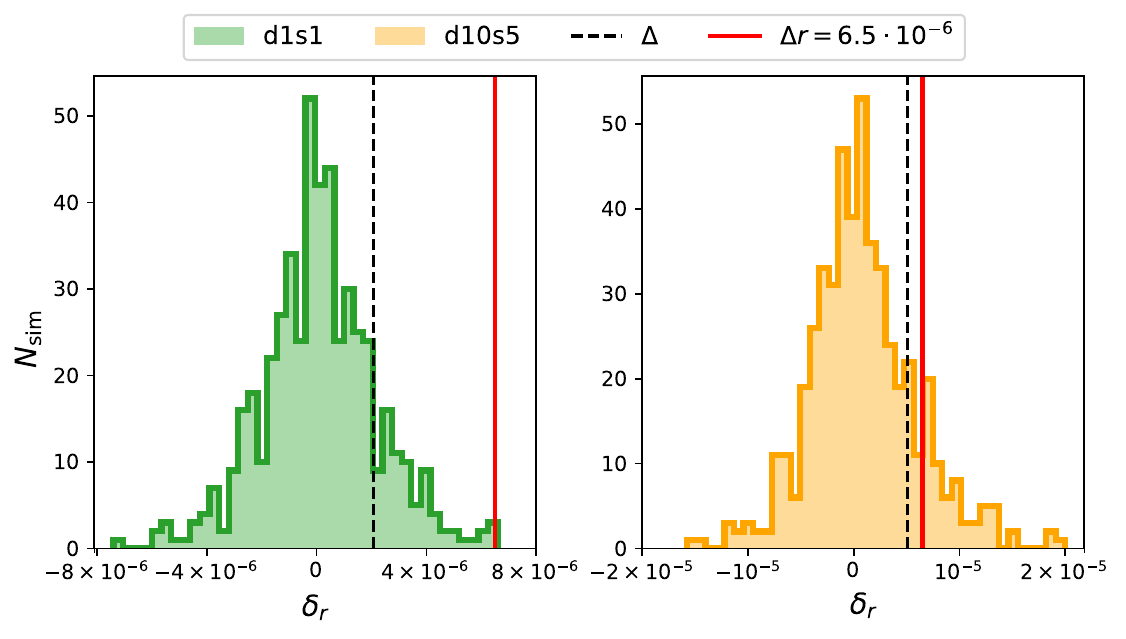}
    \caption{Distribution of ${\delta r}$ for 500 simulations when all channels are perturbed simultaneously with their corresponding requirements ${\Delta g}_{\text{req}} (\nu)$. The green histogram represents the distribution of ${\delta r}$ when the ideal MC-NILC pipeline is applied on the \textit{LiteBIRD} data set with the \texttt{d1s1} sky model while the yellow one corresponds to the \texttt{d10s5} case. The red solid line represents the budget allocated to gain systematics $\Delta r = 6.5 \times {10}^{-6}$ and the dashed black lines show the value of $\Delta = \sqrt{{\mu}^{2} + {\sigma}^{2}}$. } 
    \label{biasnewsky}
\end{figure}

For each simulation, the value of $\delta_r$ is obtained from the difference between the tensor-to-scalar ratio for the perturbed $r(\Delta g \neq 0)$  and unperturbed cases $r(\Delta g=0)$ from \cref{bias}, derived using the cosmological likelihood (eq.~\ref{Likelihood}) and considering the power spectrum of respective residuals (eq.~\ref{psres}). The distributions of  $\delta_r$ for the different sky models when the ideal MC-NILC formalism is employed are shown in \cref{biasnewsky}.

   \Cref{biasnewsky} shows that if the ideal MC-NILC component-separation technique is applied on \texttt{d1s1} data (left panel),  $\Delta$ is lower than the total gain systematics budget by a factor of $3$. Such a margin is expected, since the ideal version of MC-NILC is able to perform foreground subtraction on the \texttt{d1s1} sky almost as efficiently as the baseline NILC in a case with isotropic spectral parameters of the foregrounds. However, the slightly larger contamination by residuals in the MC-NILC \texttt{d1s1} case leads to a reduction of the margin with respect to that found in  \cref{gaincomb} with NILC and the \texttt{d0s0} model. When ideal MC-NILC is applied to the \texttt{d10s5} data set, the obtained $\Delta$ value is still below the budget, but larger than in the \texttt{d0s0} and \texttt{d1s1} cases. This is due to the higher complexity of the sky model, which leads to larger residuals in the CMB reconstruction. This observed trend of $\Delta$ highlights that the impact of {gain calibration uncertainties} on the estimation of the tensor-to-scalar ratio depends on the overall amplitude of residuals and therefore on the effectiveness of the foreground-cleaning step.
   
In a realistic framework, one way of mitigating the effect of component-separation uncertainties is to marginalise over the foreground residuals. For complex sky models and realistic MC-NILC component separation, the recovered best-fit value of the tensor-to-scalar ratio (without systematics) may indeed no longer be compatible with zero. The aim of such a marginalisation is therefore to reduce as much as possible the foreground residual bias after component separation. 
Therefore, in the cases where realistic MC-NILC is applied, we perform a marginalisation over foreground residuals to mimic the procedure of a realistic estimate of the tensor-to-scalar ratio for \textit{LiteBIRD}. Starting from the cosmological likelihood given by \cref{Likelihood}, we re-define the theoretical $BB$ power spectrum as
\begingroup
\large
\begin{equation}
{C_{\ell}}^{\text{th}} = r{C_{\ell}}^{\text{GW},r=1} +{C_{\ell}}^{\text{lensing}} + {C_{\ell}}^{\text{n}} + \gamma {C_{\ell}}^{\text{fg,res}}.
\label{marg}
\end{equation}
\endgroup
Again, the ${C_{\ell}}^{\text{fg,res}}$ and ${C_{\ell}}^{\text{n}}$ quantities correspond to a template of foregrounds and noise residuals, respectively, after component separation, obtained from averaging the MC-NILC noise and foreground residuals over 100 simulations, without systematic effects. {In this analysis, the employed foreground-residual template is not obtained through a fully realistic approach, but constructed by combining the weights with the input foreground frequency maps. However, we observe that performing the marginalisation using a foreground-residual template derived through a realistic and fully data-driven approach leads to very similar constraints ($3\%$ difference of $\Delta$ between both cases, in the \texttt{d10s5} configuration).  Such a result is in agreement which what has been found in recent studies on primordial $B$-mode reconstruction from \textit{LiteBIRD}. For this reason, we do not detail the results obtained when considering a realistic foreground residuals template and present only those coming from the ideal scenario.}
With the expression of the theoretical $B$-mode power spectrum given by  \cref{marg}, the full posterior is now 2-dimensional and defined for $r$ and $\gamma$ variables, with $\gamma$ representing the marginalisation factor: 

\begingroup
\large
\begin{equation}\label{Likelihood_marg}
-\text{ln} \mathcal{L}({C_{\ell}}^{\text{obs}}|r,\gamma) = \sum_{\ell}\frac{2\ell+1}{2} {f}_\textrm{sky} \Bigg[\frac{{C_{\ell}}^{\text{obs}}}{{C_{\ell}}^{\text{th}}(r,\gamma)} + \text{ln}({C_{\ell}}^{\text{th}}(r,\gamma)) - \frac{2\ell-1}{2\ell+1}\text{ln}({C_{\ell}}^{\text{obs}})\Bigg].
\end{equation}
\endgroup 
From the 2D log-likelihood defined in \cref{Likelihood_marg}, we build the likelihood on $r$ by marginalising over $\gamma$ values: 
\begingroup
\large
\begin{equation}\label{r-like}
\mathcal{L}(r) = \frac{\int \mathcal{L}(r,\gamma) d\gamma }{\int \mathcal{L}(r,\gamma) d\gamma dr} .
\end{equation}
\endgroup
Finally, as described in \cref{drestimation}, the best-fit value of the tensor-to-scalar ratio corresponds to the peak of the $r$ posterior distribution of \cref{r-like}.
Thereafter, we perform the marginalisation on the cases for which $r$ (without systematic effects) is not fully compatible with zero i.e.\ when the realistic MC-NILC pipeline is applied to \textit{LiteBIRD} data sets simulated assuming the \texttt{d1s1} and \texttt{d10s5} sky models. 

\begin{figure}
    \centering
    \includegraphics[width =\textwidth]{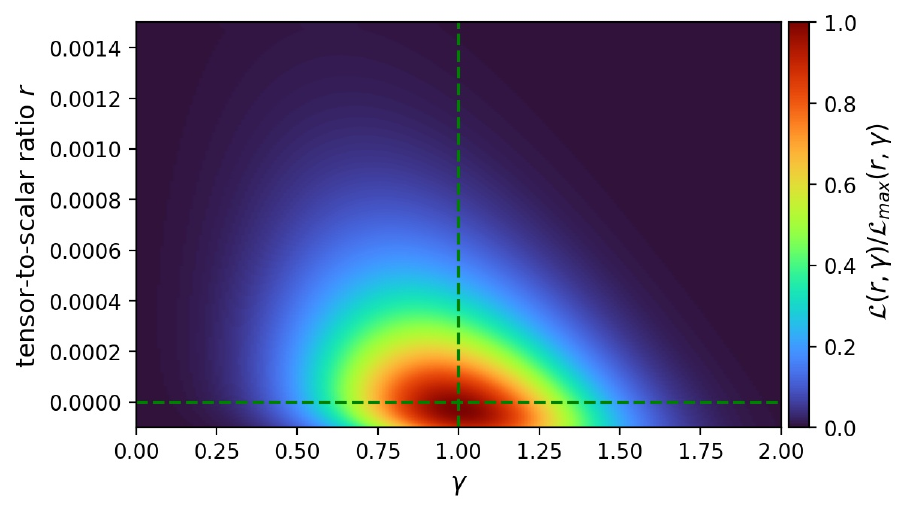}
    \caption{{Representation of the 2D likelihood $\mathcal{L}(r,\gamma)$ in the $\gamma - r$ plane. For visualisation purposes, we only display values of $r$ in the range $[-1 \times {10}^{-4},0.0015]$ and  $\gamma$ in $[0,2]$.
    The green dashed line shows the $(\gamma,r)$ pair maximising $\mathcal{L}(r,\gamma)$. }}
    \label{2d_likelihood}
\end{figure}

In practice, we retrieve the observed $BB$ residual power spectra obtained by applying the realistic MC-NILC pipeline on both \texttt{d1s1} and \texttt{d10s5} data, in the case of ideal and imperfect gain calibration. Both for ideal and mis-calibrated cases, we apply the 2D likelihood shown in \cref{Likelihood_marg} on the observed $BB$ power spectra. We consider values of $\gamma$ in the range $[0,3]$  with a step size of $0.1$, while $r$ varies in the range $[-1\times{10}^{-4},0.003]$ with a $2 \times {10}^{-7}$ step size. \Cref{2d_likelihood} represents the 2D-likelihood given by  \cref{Likelihood_marg} in the $\gamma$--$r$ plane with ${C_{\ell}}^{\text{obs}}$ (eq.~\ref{psobs}) being the average among $500$ simulated $BB$ power spectra. The ($\gamma,r$) pair that maximises the 2D-likelihood is $(1,0)$, thus demonstrating that we have an unbiased estimate of $r$. The peak value $\gamma=1$ is expected since the foreground-residual template we are marginalising over corresponds to an average among $500$ simulations and this same term appears also in the observed $BB$ power spectrum. Furthermore, we observe that $r$ and $\gamma$ are only weakly correlated. From the 2D posterior, we then estimate the value of $r$ maximising the $r$ likelihood thanks to \cref{r-like}, for both calibration cases $r^{\text{marg}}(\Delta g = 0 )$ and $r^{\text{marg}}(\Delta g \neq 0 )$, after marginalisation. Finally, we build the ${\delta r}$ distribution by differencing $r^{\text{marg}}(\Delta g = 0 )$ and $r^{\text{marg}}(\Delta g \neq 0 )$, and compare for each sky model the value of $\Delta$ with $\Delta r = 6.5 \times {10}^{-6}$. 

\begin{figure}
    \centering
    \includegraphics[width =\textwidth]{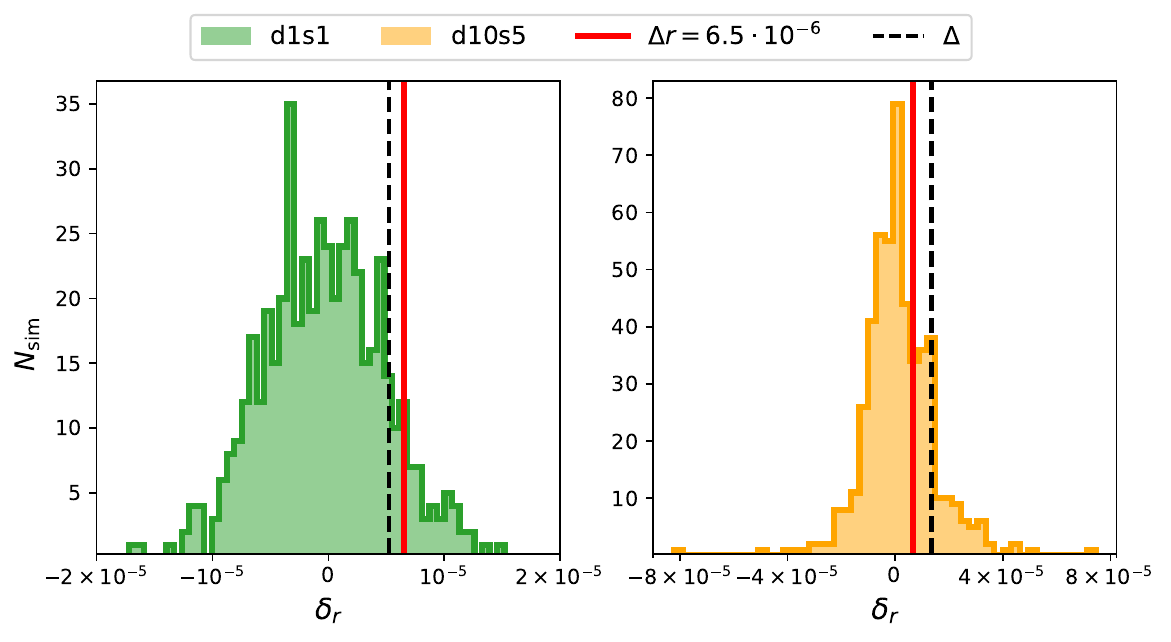}
    \caption{Distribution of ${\delta r}$ for 500 simulations when all channels are perturbed simultaneously with their corresponding requirements ${\Delta g}_{\text{req}} (\nu)$ when the realistic MC-NILC pipeline is applied on \texttt{d1s1} (left) and \texttt{d10s5} (right) \textit{LiteBIRD} simulations. The black dashed lines represent the value of $\Delta = \sqrt{{\mu}^2 + {\sigma}^2}$ of each distribution and the red solid line shows the budget allocated to gain systematics, $\Delta r = 6.5 \times {10}^{-6}$. }
    \label{Marginalisation}
\end{figure}

     \Cref{Marginalisation} shows the distribution of $\delta r$ when we apply the realistic MC-NILC pipeline to our simulations. The value of $\Delta$ obtained for \texttt{d1s1} sky model is $1.25$ times smaller than the budget while for the \texttt{d10s5} sky model $\Delta$ exceeds the threshold by a factor of approximately 2. Since this last case is the only one where we exceed the allocated budget, we now assess by what factor, denoted ${\alpha_g}^{\text{s5,d10}}$, we should reduce the requirements presented in \cref{LITEBIRDPAR} to match the total gain systematics budget. As done in \cref{all_miscal}, we interpolate the trend of $\Delta$ as a function of ${\alpha_g}^{\text{s5,d10}}$ and find the value of the factor for which $\Delta$ matches with $\Delta r = 6.5 \times {10}^{-6}$. We find that the initial requirements need to be reduced by a factor ${\alpha_g}^{\text{s5,d10}} \approx 1.8$ to fulfill the budget, in the case of the \texttt{d10s5} sky model with the realistic version of MC-NILC (after marginalisation).

{The results presented in \cref{newskymodels} highlight a clear dependence of the impact of gain calibration uncertainties on the assumed sky model. Indeed, we find that by adopting the same component-separation method (i.e.\ realistic MC-NILC), the effect induced by gain calibration uncertainties on the tensor-to-scalar ratio estimation is more significant for the \texttt{d10s5} case than for \texttt{d1s1}, with the former requiring a rescaling of the requirements derived with NILC and the \texttt{d0s0} sky model. Given our current ignorance of the true sky model and on the unpredictability of future refinements of the component-separation pipelines, it is not trivial yet to set definitive requirements without assuming a specific sky model and foreground-cleaning technique. We therefore propose a range of requirements whose limits correspond to the most optimistic and pessimistic scenarios presented in this paper.  The lower bound is given by the set of  requirements derived assuming the \texttt{d0s0} sky model with NILC  (see \cref{LITEBIRDPAR}), while the upper bound corresponds to the set of requirements rescaled by a factor of $1.8$, as needed to meet the allocated budget in the frame of the \texttt{d10s5} model with realistic MC-NILC as the component-separation method (including marginalisation). }

\section{Conclusions} \label{dis}
The space-borne mission \textit{LiteBIRD} will target a detection of the primordial tensor perturbations with an overall sensitivity of $\sigma_r \lesssim {10}^{-3}$. This requires an exquisite calibration accuracy to mitigate the impact of systematic effects, which otherwise would bias the measurement of $r$.
 
In this paper, we have presented a methodology that allows us to derive the requirements on the calibration accuracy of the relative polarisation gain, considering the application of a component-separation pipeline to reconstruct the CMB polarisation $B$-mode signal in the presence of Galactic foreground emission. Although the presented methodology is general and could be used with any component-separation technique, in our work we have made use of the NILC foreground-cleaning technique (eq.~\ref{compsep}). Unlike for parametric component-separation approaches, NILC performs a reconstruction of the CMB signal without any assumption on the Galactic foreground SED, thus representing a robust technique for any effective model of Galactic emission.

With our procedure, we first set requirements on the gain calibration accuracy needed in each frequency channel of \textit{LiteBIRD}. We did so by simulating the mis-calibration of a single frequency map, applying a homogeneous and constant gain calibration factor $g_\nu$ drawn from a Gaussian distribution centred at $1$ with standard deviation $\Delta g_\nu$, while all other channels were left unperturbed. We then derived the bias on $r$, by computing the difference $\delta_r$ between the estimated $r$ for an ideal calibration (i.e.\ $g_\nu = 1$) and in the case of an imperfect calibration. We applied a cosmological likelihood (eq.~\ref{drestimation}) on the residual $B$-mode angular power spectra after component separation (see eqs.~\ref{psres} and \ref{Likelihood}).
For each frequency, we considered different values of $\Delta g_\nu$  and determined the empirical relationship between $\Delta g_\nu$ and the quantity  $\Delta = \sqrt{{\mu}^{2}+ {\sigma}^{2}}$, with $\mu$ and $\sigma$ being the mean and standard deviation of the distribution of the bias on the tensor-to-scalar ratio $\delta_r$. This allowed us to estimate the requirement $\Delta g_{\text{req}}$ as the value for which we find $\Delta = {\delta_r}^{\text{req}} = 6.5 \times {10}^{-6} /22 $ (eq.~\ref{sumproc}).

We emphasise that a similar methodology has been applied in the past to obtain requirements for \textit{LiteBIRD}, adopting a parametric component-separation pipeline to recover the CMB signal \cite{Gainparametric}. In \cref{drdg}, we showed that the requirements on the relative polarisation gain obtained with NILC are less constraining compared to those obtained through the parametric approach \cite{Gainparametric}. Specifically, as shown in \cref{LITEBIRDPAR}, we found that the frequency channels that are more impacted by the presence of {gain calibration uncertainties} are located around the minimum of foreground emission (119, 140, and 166\,GHz), while in Ref.~\cite{Gainparametric} the most sensitive channels appear to be at low and high frequencies, where diffuse Galactic emission is dominant, and thus their mis-calibration largely affects the fit of foreground SEDs. In the case of NILC, indeed the requirements are anti-correlated with the weights (larger for central frequencies) and therefore the channels contributing the most in the CMB reconstruction are those most affected by {gain calibration uncertainties}. The interpretation of these results relies on the different approaches to foreground cleaning. While NILC absorbs the perturbation in a re-adjustment of weights into the minimum variance combination of channels, the parametric fitting requires a dedicated parametrisation to marginalise over the corresponding unknowns, in order to lower the impact of the systematic effect. In the absence of marginalisation, parametric methods, which assume a well-defined spectral dependence, can produce strongly biased CMB reconstructions.

We also simulated the mis-calibration of all channels at once, using the corresponding channel requirements. In this case, we found a $\Delta$ value lower than the total gain systematic budget by a factor of approximately $5$. Considering the available margin, we rescale the requirements at all frequency channels by a common factor $\alpha_g$: ${\Delta g}^{\prime}_{\textrm{req}}(\nu)=\alpha_g {\Delta g}_{\textrm{req}}(\nu)$ and derive that $\alpha_g\approx4.4$ allows us to match the total gain systematic budget, $\Delta r= 6.5 \times {10}^{-6}$. We highlight that, similarly to what was done in Ref.~\cite{Gainparametric}, this analysis has been performed within a simple foreground modelling frame, without any spatial variability of the polarised dust and synchrotron spectral parameters.

We therefore repeated the analysis with the injection of {gain calibration uncertainties} (according to the requirements derived for the simplest sky model) to all channels simultaneously when spatial variations of foreground spectral properties are assumed in the sky model. In order to account for these spatial variations of foregrounds spectral parameters, we used both the ideal and realistic approaches of the MC-NILC algorithm, described in \cref{newskymodels}, as the component-separation technique. We found that for the ideal version of MC-NILC,  we are able to fulfill the total gain systematics budget 
for both the \texttt{d1s1} and \texttt{d10s5} sky models ($\Delta$ below $\Delta r= 6.5 \times {10}^{-6}$ by a factor of $3$ and $1.3$, respectively).  In the frame of realistic MC-NILC and for both sky models, we can marginalise over the foreground residuals, since in these specific cases, the estimation of $r$ is distorted by the foreground-residual bias. As already commented in \cref{newskymodels}, marginalisation over foreground residuals will be one of the steps of any realistic estimate of the tensor-to-scalar ratio from \textit{LiteBIRD} data.
In the case of the application of realistic MC-NILC, we found that the value of $\Delta$ remains below the total gain systematics budget for the \texttt{d1s1} sky model, while slightly exceeding it for \texttt{d10s5} (about $2$ times larger). We thus estimated the factor ${\alpha_g}^{\text{s5,d10}}$ by which we should diminish the requirements to match the assigned gain-systematics budget when the realistic MC-NILC method is applied on \texttt{d10s5} simulations and obtained ${\alpha_g}^{\text{s5,d10}} \approx 1.8$. {Therefore, given the dependence of the gain-calibration uncertainties on the assumed sky model, we can only set a range for the requirements on gain calibration of each frequency channel. The lower bounds were found considering a simplistic (optimistic) scenario (NILC with \texttt{d0s0} foregrounds) and are reported in \cref{LITEBIRDPAR}, while the upper limits were obtained by simply rescaling the optimistic ones by a factor ${\alpha_g}^{\text{s5,d10}} \approx 1.8$, as derived in a more pessimistic case where the \texttt{d10s5} sky model is assumed.}

The effectiveness of the more realistic MC-NILC approach can be augmented by combining it with alternative (semi-)blind approaches, such as the (optimised) constrained-moment ILC (o)cMILC \cite{cMILC,ocMILC}. These methods aim at deprojecting foreground moments in order to retrieve a minimum foreground variance solution, being particularly effective on intermediate and small angular scales. Therefore, future studies may be conducted with an improved blind component-separation approach that takes advantage of both MC-NILC and (o)cMILC implementation at different angular scales.
The procedure presented in this paper is quite general and could eventually be applied to any kind of systematic effect, such as beam far sidelobes \cite{Beamsys}, for which requirements have been derived in the context of \textit{LiteBIRD}, finding very stringent requirements on the accuracy of the beam knowledge, especially at synchrotron- and dust-dominated frequencies. 

As a concluding remark, in this analysis, we assessed the impact of an imperfect relative polarisation gain calibration on the tensor-to-scalar ratio estimation, without accounting for the possible coupling of this with other types of systematic effect. For example, it is known that, the presence of a non-ideal rotating half-wave plate \cite{HWPsys} can potentially induce a mixing of the Stokes parameters (specifically, intensity-to-polarisation leakage), directly affecting gain variations and therefore our requirements. The impact of the combination of all instrumental systematic effects on component separation through an end-to-end analysis is left for future work.

\section{Acknowledgements}

This work is supported in Japan by ISAS/JAXA for Pre-Phase A2 studies, by the acceleration pro- gram of JAXA research and development directorate, by the World Premier International Research Center Initiative (WPI) of MEXT, by the JSPS Core-to-Core Program of A. Advanced Research Networks, and by JSPS KAKENHI Grant Numbers JP15H05891, JP17H01115, and JP17H01125. The Canadian contribution is supported by the Canadian Space Agency. The French LiteBIRD phase A contribution is supported by the Centre National d'Etudes Spatiale (CNES), by the Centre National de la Recherche Scientifique (CNRS), and by the Commissariat {\'a} l’Energie Atomique (CEA). The German participation in LiteBIRD is supported in part by the Excellence Cluster ORIGINS, which is funded by the Deutsche Forschungsgemeinschaft (DFG, German Research Foundation) under Germany’s Excellence Strategy (Grant No. EXC-2094--390783311). The Italian LiteBIRD phase A contribution is supported by the Italian Space Agency (ASI Grants No.~2020-9-HH.0 and 2016-24- H.1-2018), the National Institute for Nuclear Physics (INFN) and the National Institute for Astrophysics (INAF). Norwegian participation in LiteBIRD is supported by the Research Council of Norway (Grant No.~263011) and has received funding from the European Research Council (ERC) under the Horizon 2020 Research and Innovation Programme (Grant agreement No.~772253 and 819478). The Spanish LiteBIRD phase A contribution is supported by the Spanish Agencia Estatal de Investigaci{\'o}n (AEI), project refs.~PID2019-110610RB-C21, PID2020-120514GB-I00, ProID2020010108 and ICTP20210008. Funds that support contributions from Sweden come from the Swedish National Space Agency (SNSA/Rymdstyrelsen) and the Swedish Research Council (Reg.\ no.\ 2019-03959). The US contribution is supported by NASA grant no.~80NSSC18K0132.
This work has also received funding by the European Union’s Horizon 2020 research and innovation programme under grant agreement no.~101007633 CMB-Inflate. It also received partial support from the Italian Space Agency LiteBIRD Project (ASI Grants No. 2020-9-HH.0 and 2016-24-H.1-2018), as well as the LiteBIRD Initiative of the National Institute for Nuclear Phyiscs, and the RadioForegroundsPlus Project HORIZON-CL4-2023-SPACE-01, GA 101135036. 
This work benefited from computational resources provided by the National Energy Research Scientific Computing Center (NERSC), managed by the Lawrence Berkeley National Laboratory for U.S. Department of Energy.

\bibliographystyle{JHEP}
\bibliography{bibliographie}

\providecommand{\href}[2]{#2}\begingroup\raggedright\begin{thebibliography}{10}

\bibitem{CMB}
A.A.~{Penzias} and R.W.~{Wilson}, \emph{{A Measurement of Excess Antenna Temperature at 4080 Mc/s.}}, \href{https://doi.org/10.1086/148307}{\emph{\apj} {\bfseries 142} (1965) 419}.

\bibitem{COBE}
C.H.~{Lineweaver}, L.~{Tenorio}, G.F.~{Smoot}, P.~{Keegstra}, A.J.~{Banday} and P.~{Lubin}, \emph{{The Dipole Observed in the COBE DMR 4 Year Data}}, \href{https://doi.org/10.1086/177846}{\emph{\apj} {\bfseries 470} (1996) 38} [\href{https://arxiv.org/abs/astro-ph/9601151}{{\ttfamily astro-ph/9601151}}].

\bibitem{WMAP}
C.L.~{Bennett}, D.~{Larson}, J.L.~{Weiland}, N.~{Jarosik}, G.~{Hinshaw}, N.~{Odegard} et~al., \emph{{Nine-year Wilkinson Microwave Anisotropy Probe (WMAP) Observations: Final Maps and Results}}, \href{https://doi.org/10.1088/0067-0049/208/2/20}{\emph{\apjs} {\bfseries 208} (2013) 20} [\href{https://arxiv.org/abs/1212.5225}{{\ttfamily 1212.5225}}].

\bibitem{Planck2013}
{\scshape Planck} collaboration, \emph{{Planck 2013 results. I. Overview of products and scientific results}}, \href{https://doi.org/10.1051/0004-6361/201321529}{\emph{\aap} {\bfseries 571} (2014) A1} [\href{https://arxiv.org/abs/1303.5062}{{\ttfamily 1303.5062}}].

\bibitem{Planck2015}
{\scshape Planck} collaboration, \emph{{Planck 2015 results. I. Overview of products and scientific results}}, \href{https://doi.org/10.1051/0004-6361/201527101}{\emph{\aap} {\bfseries 594} (2016) A1} [\href{https://arxiv.org/abs/1502.01582}{{\ttfamily 1502.01582}}].

\bibitem{Planck2018}
{\scshape Planck} collaboration, \emph{{Planck 2018 results. I. Overview and the cosmological legacy of Planck}}, \href{https://doi.org/10.1051/0004-6361/201833880}{\emph{\aap} {\bfseries 641} (2020) A1} [\href{https://arxiv.org/abs/1807.06205}{{\ttfamily 1807.06205}}].

\bibitem{boomerang}
P.~{de Bernardis}, P.A.R.~{Ade}, J.J.~{Bock}, J.R.~{Bond}, J.~{Borrill}, A.~{Boscaleri} et~al., \emph{{First results from the BOOMERanG experiment}},  in \emph{Cosmology and Particle Physics}, R.~{Durrer}, J.~{Garcia-Bellido} and M.~{Shaposhnikov}, eds., vol.~555 of \emph{American Institute of Physics Conference Series}, pp.~85--94, AIP, Feb., 2001, \href{https://doi.org/10.1063/1.1363510}{DOI} [\href{https://arxiv.org/abs/astro-ph/0011469}{{\ttfamily astro-ph/0011469}}].

\bibitem{Balkenhol_2023}
L.~{Balkenhol}, D.~{Dutcher}, A.~{Spurio Mancini}, A.~{Doussot}, K.~{Benabed}, S.~{Galli} et~al., \emph{{Measurement of the CMB temperature power spectrum and constraints on cosmology from the SPT-3G 2018 $TT$, $TE$, and $EE$ dataset}}, \href{https://doi.org/10.1103/PhysRevD.108.023510}{\emph{\prd} {\bfseries 108} (2023) 023510} [\href{https://arxiv.org/abs/2212.05642}{{\ttfamily 2212.05642}}].

\bibitem{madhavacheril2023atacama}
M.S.~{Madhavacheril}, F.J.~{Qu}, B.D.~{Sherwin}, N.~{MacCrann}, Y.~{Li}, I.~{Abril-Cabezas} et~al., \emph{{The Atacama Cosmology Telescope: DR6 Gravitational Lensing Map and Cosmological Parameters}}, \href{https://doi.org/10.3847/1538-4357/acff5f}{\emph{\apj} {\bfseries 962} (2024) 113} [\href{https://arxiv.org/abs/2304.05203}{{\ttfamily 2304.05203}}].

\bibitem{inf}
A.H.~{Guth}, \emph{{Inflationary universe: A possible solution to the horizon and flatness problems}}, \href{https://doi.org/10.1103/PhysRevD.23.347}{\emph{\prd} {\bfseries 23} (1981) 347}.

\bibitem{1997kamion}
M.~{Kamionkowski}, A.~{Kosowsky} and A.~{Stebbins}, \emph{{A Probe of Primordial Gravity Waves and Vorticity}}, \href{https://doi.org/10.1103/PhysRevLett.78.2058}{\emph{Physical Review Letters} {\bfseries 78} (1997) 2058} [\href{https://arxiv.org/abs/astro-ph/9609132}{{\ttfamily astro-ph/9609132}}].

\bibitem{HuWhite}
W.~{Hu} and M.~{White}, \emph{{A CMB polarization primer}}, \href{https://doi.org/10.1016/S1384-1076(97)00022-5}{\emph{New Astronomy} {\bfseries 2} (1997) 323} [\href{https://arxiv.org/abs/astro-ph/9706147}{{\ttfamily astro-ph/9706147}}].

\bibitem{SeljakZaldarriaga}
U.~{Seljak} and M.~{Zaldarriaga}, \emph{{Signature of Gravity Waves in the Polarization of the Microwave Background}}, \href{https://doi.org/10.1103/PhysRevLett.78.2054}{\emph{\prl} {\bfseries 78} (1997) 2054} [\href{https://arxiv.org/abs/astro-ph/9609169}{{\ttfamily astro-ph/9609169}}].

\bibitem{infplanck}
{\scshape Planck} collaboration, \emph{{Planck 2018 results. X. Constraints on inflation}}, \href{https://doi.org/10.1051/0004-6361/201833887}{\emph{\aap} {\bfseries 641} (2020) A10} [\href{https://arxiv.org/abs/1807.06211}{{\ttfamily 1807.06211}}].

\bibitem{tristram2022}
M.~{Tristram}, A.J.~{Banday}, K.M.~{G{\'o}rski}, R.~{Keskitalo}, C.R.~{Lawrence}, K.J.~{Andersen} et~al., \emph{{Improved limits on the tensor-to-scalar ratio using BICEP and Planck data}}, \href{https://doi.org/10.1103/PhysRevD.105.083524}{\emph{\prd} {\bfseries 105} (2022) 083524} [\href{https://arxiv.org/abs/2112.07961}{{\ttfamily 2112.07961}}].

\bibitem{galloni2023}
G.~{Galloni}, N.~{Bartolo}, S.~{Matarrese}, M.~{Migliaccio}, A.~{Ricciardone} and N.~{Vittorio}, \emph{{Updated constraints on amplitude and tilt of the tensor primordial spectrum}}, \href{https://doi.org/10.1088/1475-7516/2023/04/062}{\emph{\jcap} {\bfseries 2023} (2023) 062} [\href{https://arxiv.org/abs/2208.00188}{{\ttfamily 2208.00188}}].

\bibitem{polarbear}
{\scshape Polarbear} collaboration, \emph{{Measurement of the Cosmic Microwave Background Polarization Lensing Power Spectrum with the POLARBEAR Experiment}}, \href{https://doi.org/10.1103/PhysRevLett.113.021301}{\emph{\prl} {\bfseries 113} (2014) 021301} [\href{https://arxiv.org/abs/1312.6646}{{\ttfamily 1312.6646}}].

\bibitem{bicep}
{\scshape BICEP/Keck} collaboration, \emph{{Measurements of Degree-Scale B-mode Polarization with the BICEP/Keck Experiments at South Pole}}, \href{https://doi.org/10.48550/arXiv.1807.02199}{\emph{arXiv e-prints} (2018) } [\href{https://arxiv.org/abs/1807.02199}{{\ttfamily 1807.02199}}].

\bibitem{Lensing}
A.~{Blanchard} and J.~{Schneider}, \emph{{Gravitational lensing effect on the fluctuations of the cosmic background radiation}}, {\emph{\aap} {\bfseries 184} (1987) [\href{https://ui.adsabs.harvard.edu/abs/1987A&A...184....1B}{1987A\&A...184....1B}]}.

\bibitem{Lensing2}
M.~{Zaldarriaga} and U.~{Seljak}, \emph{{Gravitational lensing effect on cosmic microwave background polarization}}, \href{https://doi.org/10.1103/PhysRevD.58.023003}{\emph{\prd} {\bfseries 58} (1998) 023003} [\href{https://arxiv.org/abs/astro-ph/9803150}{{\ttfamily astro-ph/9803150}}].

\bibitem{SPTpol}
E.J.~{Baxter}, R.~{Keisler}, S.~{Dodelson}, K.A.~{Aird}, S.W.~{Allen}, M.L.N.~{Ashby} et~al., \emph{{A Measurement of Gravitational Lensing of the Cosmic Microwave Background by Galaxy Clusters Using Data from the South Pole Telescope}}, \href{https://doi.org/10.1088/0004-637X/806/2/247}{\emph{\apj} {\bfseries 806} (2015) 247} [\href{https://arxiv.org/abs/1412.7521}{{\ttfamily 1412.7521}}].

\bibitem{ACTPOL}
B.D.~{Sherwin}, A.~{van Engelen}, N.~{Sehgal}, M.~{Madhavacheril}, G.E.~{Addison}, S.~{Aiola} et~al., \emph{{Two-season Atacama Cosmology Telescope polarimeter lensing power spectrum}}, \href{https://doi.org/10.1103/PhysRevD.95.123529}{\emph{\prd} {\bfseries 95} (2017) 123529} [\href{https://arxiv.org/abs/1611.09753}{{\ttfamily 1611.09753}}].

\bibitem{Fg}
N.~{Krachmalnicoff}, C.~{Baccigalupi}, J.~{Aumont}, M.~{Bersanelli} and A.~{Mennella}, \emph{{Characterization of foreground emission on degree angular scales for CMB B-mode observations . Thermal dust and synchrotron signal from Planck and WMAP data}}, \href{https://doi.org/10.1051/0004-6361/201527678}{\emph{\aap} {\bfseries 588} (2016) A65} [\href{https://arxiv.org/abs/1511.00532}{{\ttfamily 1511.00532}}].

\bibitem{Rybicki}
G.B.~Rybicki and A.P.~Lightman, \emph{{Radiative Processes in Astrophysics}}, Wiley, New York, NY (1985), \href{https://doi.org/10.1002/9783527618170}{10.1002/9783527618170}.

\bibitem{2011MNRAS.418..888M}
N.~{Macellari}, E.~{Pierpaoli}, C.~{Dickinson} and J.E.~{Vaillancourt}, \emph{{Galactic foreground contributions to the 5-year Wilkinson Microwave Anisotropy Probe maps}}, \href{https://doi.org/10.1111/j.1365-2966.2011.19542.x}{\emph{\mnras} {\bfseries 418} (2011) 888} [\href{https://arxiv.org/abs/1108.0205}{{\ttfamily 1108.0205}}].

\bibitem{2011ame}
C.~{Dickinson}, M.~{Peel} and M.~{Vidal}, \emph{{New constraints on the polarization of anomalous microwave emission in nearby molecular clouds}}, \href{https://doi.org/10.1111/j.1745-3933.2011.01138.x}{\emph{\mnras} {\bfseries 418} (2011) L35} [\href{https://arxiv.org/abs/1108.0308}{{\ttfamily 1108.0308}}].

\bibitem{AME_commander}
D.~{Herman}, B.~{Hensley}, K.J.~{Andersen}, R.~{Aurlien}, R.~{Banerji}, M.~{Bersanelli} et~al., \emph{{BEYONDPLANCK. XV. Limits on large-scale polarized anomalous microwave emission from Planck LFI and WMAP}}, \href{https://doi.org/10.1051/0004-6361/202243081}{\emph{\aap} {\bfseries 675} (2023) A15} [\href{https://arxiv.org/abs/2201.03530}{{\ttfamily 2201.03530}}].

\bibitem{2017MNRAS.464.4107G}
R.~{G{\'e}nova-Santos}, J.A.~{Rubi{\~n}o-Mart{\'\i}n}, A.~{Pel{\'a}ez-Santos}, F.~{Poidevin}, R.~{Rebolo}, R.~{Vignaga} et~al., \emph{{QUIJOTE scientific results - II. Polarisation measurements of the microwave emission in the Galactic molecular complexes W43 and W47 and supernova remnant W44}}, \href{https://doi.org/10.1093/mnras/stw2503}{\emph{\mnras} {\bfseries 464} (2017) 4107} [\href{https://arxiv.org/abs/1605.04741}{{\ttfamily 1605.04741}}].

\bibitem{ame_quijote}
R.~{Gonz{\'a}lez-Gonz{\'a}lez}, R.T.~{G{\'e}nova-Santos}, J.A.~{Rubi{\~n}o-Mart{\'\i}n}, M.W.~{Peel}, F.~{Guidi}, C.H.~{L{\'o}pez-Caraballo} et~al., \emph{{QUIJOTE scientific results -- XVIII. New constraints on the polarization of the Anomalous Microwave Emission in bright Galactic regions: $\rho$\,Ophiuchi, Perseus and W43}}, \href{https://doi.org/10.48550/arXiv.2409.03418}{\emph{arXiv e-prints} (2024) } [\href{https://arxiv.org/abs/2409.03418}{{\ttfamily 2409.03418}}].

\bibitem{CO}
G.~{Puglisi}, G.~{Fabbian} and C.~{Baccigalupi}, \emph{{A 3D model for carbon monoxide molecular line emission as a potential cosmic microwave background polarization contaminant}}, \href{https://doi.org/10.1093/mnras/stx1029}{\emph{\mnras} {\bfseries 469} (2017) 2982} [\href{https://arxiv.org/abs/1701.07856}{{\ttfamily 1701.07856}}].

\bibitem{dustplanck}
{\scshape Planck} collaboration, \emph{{Planck 2018 results. XI. Polarized dust foregrounds}}, \href{https://doi.org/10.1051/0004-6361/201832618}{\emph{\aap} {\bfseries 641} (2020) A11} [\href{https://arxiv.org/abs/1801.04945}{{\ttfamily 1801.04945}}].

\bibitem{dustcomp}
V.~{Pelgrims}, S.E.~{Clark}, B.S.~{Hensley}, G.V.~{Panopoulou}, V.~{Pavlidou}, K.~{Tassis} et~al., \emph{{Evidence for line-of-sight frequency decorrelation of polarized dust emission in Planck data}}, \href{https://doi.org/10.1051/0004-6361/202040218}{\emph{\aap} {\bfseries 647} (2021) A16} [\href{https://arxiv.org/abs/2101.09291}{{\ttfamily 2101.09291}}].

\bibitem{Skalidis2018}
R.~{Skalidis}, G.V.~{Panopoulou}, K.~{Tassis}, V.~{Pavlidou}, D.~{Blinov}, I.~{Komis} et~al., \emph{{Local measurements of the mean interstellar polarization at high Galactic latitudes}}, \href{https://doi.org/10.1051/0004-6361/201832827}{\emph{\aap} {\bfseries 616} (2018) A52} [\href{https://arxiv.org/abs/1802.04305}{{\ttfamily 1802.04305}}].

\bibitem{compsepplanck}
{\scshape Planck} collaboration, \emph{{Planck 2018 results. IV. Diffuse component separation}}, \href{https://doi.org/10.1051/0004-6361/201833881}{\emph{\aap} {\bfseries 641} (2020) A4} [\href{https://arxiv.org/abs/1807.06208}{{\ttfamily 1807.06208}}].

\bibitem{Commander}
H.K.~{Eriksen}, J.B.~{Jewell}, C.~{Dickinson}, A.J.~{Banday}, K.M.~{G{\'o}rski} and C.R.~{Lawrence}, \emph{{Joint Bayesian Component Separation and CMB Power Spectrum Estimation}}, \href{https://doi.org/10.1086/525277}{\emph{\apj} {\bfseries 676} (2008) 10} [\href{https://arxiv.org/abs/0709.1058}{{\ttfamily 0709.1058}}].

\bibitem{param}
R.~{Stompor}, S.~{Leach}, F.~{Stivoli} and C.~{Baccigalupi}, \emph{{Maximum likelihood algorithm for parametric component separation in cosmic microwave background experiments}}, \href{https://doi.org/10.1111/j.1365-2966.2008.14023.x}{\emph{\mnras} {\bfseries 392} (2009) 216} [\href{https://arxiv.org/abs/0804.2645}{{\ttfamily 0804.2645}}].

\bibitem{Bsecret}
E.~{De la Hoz}, P.~{Vielva}, R.B.~{Barreiro} and E.~{Mart{\'\i}nez-Gonz{\'a}lez}, \emph{{On the detection of CMB B-modes from ground at low frequency}}, \href{https://doi.org/10.1088/1475-7516/2020/06/006}{\emph{\jcap} {\bfseries 2020} (2020) 006} [\href{https://arxiv.org/abs/2002.12206}{{\ttfamily 2002.12206}}].

\bibitem{Azzoni2021}
S.~{Azzoni}, M.H.~{Abitbol}, D.~{Alonso}, A.~{Gough}, N.~{Katayama} and T.~{Matsumura}, \emph{{A minimal power-spectrum-based moment expansion for CMB B-mode searches}}, \href{https://doi.org/10.1088/1475-7516/2021/05/047}{\emph{\jcap} {\bfseries 2021} (2021) 047} [\href{https://arxiv.org/abs/2011.11575}{{\ttfamily 2011.11575}}].

\bibitem{Vacher2022}
L.~{Vacher}, J.~{Aumont}, L.~{Montier}, S.~{Azzoni}, F.~{Boulanger} and M.~{Remazeilles}, \emph{{Moment expansion of polarized dust SED: A new path towards capturing the CMB B-modes with LiteBIRD}}, \href{https://doi.org/10.1051/0004-6361/202142664}{\emph{\aap} {\bfseries 660} (2022) A111} [\href{https://arxiv.org/abs/2111.07742}{{\ttfamily 2111.07742}}].

\bibitem{Commander3}
M.~{Galloway}, K.J.~{Andersen}, R.~{Aurlien}, R.~{Banerji}, M.~{Bersanelli}, S.~{Bertocco} et~al., \emph{{BEYONDPLANCK. III. Commander3}}, \href{https://doi.org/10.1051/0004-6361/202243137}{\emph{\aap} {\bfseries 675} (2023) A3} [\href{https://arxiv.org/abs/2201.03509}{{\ttfamily 2201.03509}}].

\bibitem{ILCWMAP}
J.~{Delabrouille}, J.F.~{Cardoso}, M.~{Le Jeune}, M.~{Betoule}, G.~{Fay} and F.~{Guilloux}, \emph{{A full sky, low foreground, high resolution CMB map from WMAP}}, \href{https://doi.org/10.1051/0004-6361:200810514}{\emph{\aap} {\bfseries 493} (2009) 835} [\href{https://arxiv.org/abs/0807.0773}{{\ttfamily 0807.0773}}].

\bibitem{ILCharmonic}
R.~{Vio} and P.~{Andreani}, \emph{{``Internal Linear Combination'' method for the separation of CMB from Galactic foregrounds in the harmonic domain}}, \href{https://doi.org/10.48550/arXiv.0811.4277}{\emph{arXiv e-prints} (2008) } [\href{https://arxiv.org/abs/0811.4277}{{\ttfamily 0811.4277}}].

\bibitem{genILC}
M.~{Remazeilles}, J.~{Delabrouille} and J.-F.~{Cardoso}, \emph{{Foreground component separation with generalized Internal Linear Combination}}, \href{https://doi.org/10.1111/j.1365-2966.2011.19497.x}{\emph{\mnras} {\bfseries 418} (2011) 467} [\href{https://arxiv.org/abs/1103.1166}{{\ttfamily 1103.1166}}].

\bibitem{constrainedILC}
M.~{Remazeilles}, J.~{Delabrouille} and J.-F.~{Cardoso}, \emph{{CMB and SZ effect separation with constrained Internal Linear Combinations}}, \href{https://doi.org/10.1111/j.1365-2966.2010.17624.x}{\emph{\mnras} {\bfseries 410} (2011) 2481} [\href{https://arxiv.org/abs/1006.5599}{{\ttfamily 1006.5599}}].

\bibitem{NILC_dela}
J.~{Delabrouille}, J.-F.~{Cardoso}, M.~{Le Jeune}, M.~{Betoule}, G.~{Fay} and F.~{Guilloux}, \emph{{A full sky, low foreground, high resolution CMB map from WMAP}}, \href{https://doi.org/10.1051/0004-6361:200810514}{\emph{\aap} {\bfseries 493} (2009) 835} [\href{https://arxiv.org/abs/0807.0773}{{\ttfamily 0807.0773}}].

\bibitem{MCNILC}
A.~{Carones}, M.~{Migliaccio}, G.~{Puglisi}, C.~{Baccigalupi}, D.~{Marinucci}, N.~{Vittorio} et~al., \emph{{Multiclustering needlet ILC for CMB B-mode component separation}}, \href{https://doi.org/10.1093/mnras/stad2423}{\emph{\mnras} {\bfseries 525} (2023) 3117} [\href{https://arxiv.org/abs/2212.04456}{{\ttfamily 2212.04456}}].

\bibitem{Tegmark_2003}
M.~{Tegmark}, A.~{de Oliveira-Costa} and A.J.~{Hamilton}, \emph{{High resolution foreground cleaned CMB map from WMAP}}, \href{https://doi.org/10.1103/PhysRevD.68.123523}{\emph{\prd} {\bfseries 68} (2003) 123523} [\href{https://arxiv.org/abs/astro-ph/0302496}{{\ttfamily astro-ph/0302496}}].

\bibitem{LiteBIRD}
M.~{Hazumi}, P.A.R.~{Ade}, Y.~{Akiba}, D.~{Alonso}, K.~{Arnold}, J.~{Aumont} et~al., \emph{{LiteBIRD: A Satellite for the Studies of B-Mode Polarization and Inflation from Cosmic Background Radiation Detection}}, \href{https://doi.org/10.1007/s10909-019-02150-5}{\emph{\jltp} {\bfseries 194} (2019) 443}.

\bibitem{PTEP}
{\scshape LiteBIRD} collaboration, \emph{{Probing cosmic inflation with the LiteBIRD cosmic microwave background polarization survey}}, \href{https://doi.org/10.1093/ptep/ptac150}{\emph{\ptep} {\bfseries 2023} (2023) 042F01} [\href{https://arxiv.org/abs/2202.02773}{{\ttfamily 2202.02773}}].

\bibitem{spie2024}
T.~{Ghigna}, A.~{Adler}, K.~{Aizawa}, H.~{Akamatsu}, R.~{Akizawa}, E.~{Allys} et~al., \emph{{The LiteBIRD mission to explore cosmic inflation}}, \href{https://doi.org/10.48550/arXiv.2406.02724}{\emph{arXiv e-prints} (2024) arXiv:2406.02724} [\href{https://arxiv.org/abs/2406.02724}{{\ttfamily 2406.02724}}].

\bibitem{NILC}
S.~{Basak} and J.~{Delabrouille}, \emph{{A needlet ILC analysis of WMAP 9-year polarization data: CMB polarization power spectra}}, \href{https://doi.org/10.1093/mnras/stt1158}{\emph{\mnras} {\bfseries 435} (2013) 18} [\href{https://arxiv.org/abs/1204.0292}{{\ttfamily 1204.0292}}].

\bibitem{NILC2}
A.~{Carones}, M.~{Migliaccio}, D.~{Marinucci} and N.~{Vittorio}, \emph{{Analysis of Needlet Internal Linear Combination performance on B-mode data from sub-orbital experiments}}, \href{https://doi.org/10.1051/0004-6361/202244824}{\emph{\aap} {\bfseries 677} (2023) A147} [\href{https://arxiv.org/abs/2208.12059}{{\ttfamily 2208.12059}}].

\bibitem{Gainparametric}
T.~{Ghigna}, T.~{Matsumura}, G.~{Patanchon}, H.~{Ishino} and M.~{Hazumi}, \emph{{Requirements for future CMB satellite missions: photometric and band-pass response calibration}}, \href{https://doi.org/10.1088/1475-7516/2020/11/030}{\emph{\jcap} {\bfseries 2020} (2020) 030} [\href{https://arxiv.org/abs/2004.11601}{{\ttfamily 2004.11601}}].

\bibitem{Commandergain}
E.~{Gjerl{\o}w}, H.T.~{Ihle}, S.~{Galeotta}, K.J.~{Andersen}, R.~{Aurlien}, R.~{Banerji} et~al., \emph{{BEYONDPLANCK. VII. Bayesian estimation of gain and absolute calibration for cosmic microwave background experiments}}, \href{https://doi.org/10.1051/0004-6361/202244061}{\emph{\aap} {\bfseries 675} (2023) A7} [\href{https://arxiv.org/abs/2011.08082}{{\ttfamily 2011.08082}}].

\bibitem{LFT}
Y.~{Sekimoto}, P.A.R.~{Ade}, A.~{Adler}, E.~{Allys}, K.~{Arnold}, D.~{Auguste} et~al., \emph{{Concept Design of Low Frequency Telescope for CMB B-mode Polarization satellite LiteBIRD}}, \href{https://doi.org/10.48550/arXiv.2101.06342}{\emph{arXiv e-prints} (2021) arXiv:2101.06342} [\href{https://arxiv.org/abs/2101.06342}{{\ttfamily 2101.06342}}].

\bibitem{MHFT}
L.~{Montier}, B.~{Mot}, P.~{de Bernardis}, B.~{Maffei}, G.~{Pisano}, F.~{Columbro} et~al., \emph{{Overview of the medium and high frequency telescopes of the LiteBIRD space mission}},  in \emph{Space Telescopes and Instrumentation 2020: Optical, Infrared, and Millimeter Wave}, vol.~11443 of \emph{Society of Photo-Optical Instrumentation Engineers (SPIE) Conference Series}, p.~114432G, Dec., 2020, \href{https://doi.org/10.1117/12.2562243}{DOI} [\href{https://arxiv.org/abs/2102.00809}{{\ttfamily 2102.00809}}].

\bibitem{Giardiello2022}
S.~{Giardiello}, M.~{Gerbino}, L.~{Pagano}, J.~{Errard}, A.~{Gruppuso}, H.~{Ishino} et~al., \emph{{Detailed study of HWP non-idealities and their impact on future measurements of CMB polarization anisotropies from space}}, \href{https://doi.org/10.1051/0004-6361/202141619}{\emph{\aap} {\bfseries 658} (2022) A15} [\href{https://arxiv.org/abs/2106.08031}{{\ttfamily 2106.08031}}].

\bibitem{Monelli2023}
M.~{Monelli}, E.~{Komatsu}, T.~{Ghigna}, T.~{Matsumura}, G.~{Pisano} and R.~{Takaku}, \emph{{Impact of half-wave plate systematics on the measurement of CMB B-mode polarization}}, \href{https://doi.org/10.1088/1475-7516/2024/05/018}{\emph{\jcap} {\bfseries 2024} (2024) 018} [\href{https://arxiv.org/abs/2311.07999}{{\ttfamily 2311.07999}}].

\bibitem{2023arXiv230800967P}
G.~{Patanchon}, H.~{Imada}, H.~{Ishino} and T.~{Matsumura}, \emph{{Effect of instrumental polarization with a half-wave plate on the B-mode signal: prediction and correction}}, \href{https://doi.org/10.1088/1475-7516/2024/04/074}{\emph{\jcap} {\bfseries 2024} (2024) 074} [\href{https://arxiv.org/abs/2308.00967}{{\ttfamily 2308.00967}}].

\bibitem{foreg}
A.~{Kogut}, J.~{Dunkley}, C.L.~{Bennett}, O.~{Dor{\'e}}, B.~{Gold}, M.~{Halpern} et~al., \emph{{Three-Year Wilkinson Microwave Anisotropy Probe (WMAP) Observations: Foreground Polarization}}, \href{https://doi.org/10.1086/519754}{\emph{\apj} {\bfseries 665} (2007) 355} [\href{https://arxiv.org/abs/0704.3991}{{\ttfamily 0704.3991}}].

\bibitem{synchspec}
M.A.~{Miville-Desch{\^e}nes}, N.~{Ysard}, A.~{Lavabre}, N.~{Ponthieu}, J.F.~{Mac{\'\i}as-P{\'e}rez}, J.~{Aumont} et~al., \emph{{Separation of anomalous and synchrotron emissions using WMAP polarization data}}, \href{https://doi.org/10.1051/0004-6361:200809484}{\emph{\aap} {\bfseries 490} (2008) 1093} [\href{https://arxiv.org/abs/0802.3345}{{\ttfamily 0802.3345}}].

\bibitem{spec}
U.~{Fuskeland}, I.K.~{Wehus}, H.K.~{Eriksen} and S.K.~{N{\ae}ss}, \emph{{Spatial Variations in the Spectral Index of Polarized Synchrotron Emission in the 9 yr WMAP Sky Maps}}, \href{https://doi.org/10.1088/0004-637X/790/2/104}{\emph{\apj} {\bfseries 790} (2014) 104} [\href{https://arxiv.org/abs/1404.5323}{{\ttfamily 1404.5323}}].

\bibitem{SPASS}
N.~{Krachmalnicoff}, E.~{Carretti}, C.~{Baccigalupi}, G.~{Bernardi}, S.~{Brown}, B.M.~{Gaensler} et~al., \emph{{S-PASS view of polarized Galactic synchrotron at 2.3 GHz as a contaminant to CMB observations}}, \href{https://doi.org/10.1051/0004-6361/201832768}{\emph{\aap} {\bfseries 618} (2018) A166} [\href{https://arxiv.org/abs/1802.01145}{{\ttfamily 1802.01145}}].

\bibitem{PlanckPS}
{\scshape Planck} collaboration, \emph{{Planck 2018 results. V. CMB power spectra and likelihoods}}, \href{https://doi.org/10.1051/0004-6361/201936386}{\emph{\aap} {\bfseries 641} (2020) A5} [\href{https://arxiv.org/abs/1907.12875}{{\ttfamily 1907.12875}}].

\bibitem{CAMBpaper}
A.~Lewis, A.~Challinor and A.~Lasenby, \emph{{Efficient computation of CMB anisotropies in closed FRW models}}, \href{https://doi.org/10.1086/309179}{\emph{\apj} {\bfseries 538} (2000) 473} [\href{https://arxiv.org/abs/astro-ph/9911177}{{\ttfamily astro-ph/9911177}}].

\bibitem{Zaldarriaga_1997}
M.~{Zaldarriaga} and U.~{Seljak}, \emph{{All-sky analysis of polarization in the microwave background}}, \href{https://doi.org/10.1103/PhysRevD.55.1830}{\emph{\prd} {\bfseries 55} (1997) 1830} [\href{https://arxiv.org/abs/astro-ph/9609170}{{\ttfamily astro-ph/9609170}}].

\bibitem{SO}
P.~{Ade}, J.~{Aguirre}, Z.~{Ahmed}, S.~{Aiola}, A.~{Ali}, D.~{Alonso} et~al., \emph{{The Simons Observatory: science goals and forecasts}}, \href{https://doi.org/10.1088/1475-7516/2019/02/056}{\emph{\jcap} {\bfseries 2019} (2019) 056} [\href{https://arxiv.org/abs/1808.07445}{{\ttfamily 1808.07445}}].

\bibitem{ILCcalib}
J.~{Dick}, M.~{Remazeilles} and J.~{Delabrouille}, \emph{{Impact of calibration errors on CMB component separation using FastICA and ILC}}, \href{https://doi.org/10.1111/j.1365-2966.2009.15798.x}{\emph{\mnras} {\bfseries 401} (2010) 1602} [\href{https://arxiv.org/abs/0907.3105}{{\ttfamily 0907.3105}}].

\bibitem{ILC}
C.L.~{Bennett}, R.S.~{Hill}, G.~{Hinshaw}, M.R.~{Nolta}, N.~{Odegard}, L.~{Page} et~al., \emph{{First-Year Wilkinson Microwave Anisotropy Probe (WMAP) Observations: Foreground Emission}}, \href{https://doi.org/10.1086/377252}{\emph{\apjs} {\bfseries 148} (2003) 97} [\href{https://arxiv.org/abs/astro-ph/0302208}{{\ttfamily astro-ph/0302208}}].

\bibitem{HILC}
M.~{Tegmark}, A.~{de Oliveira-Costa} and A.~{Hamilton}, \emph{{High resolution foreground cleaned CMB map from WMAP}}, \href{https://doi.org/10.1103/PhysRevD.68.123523}{\emph{\prd} {\bfseries 68} (2003) 123523} [\href{https://arxiv.org/abs/astro-ph/0302496}{{\ttfamily astro-ph/0302496}}].

\bibitem{doi:10.1137/040614359}
F.J.~Narcowich, P.~Petrushev and J.D.~Ward, \emph{Localized tight frames on spheres}, \href{https://doi.org/10.1137/040614359}{\emph{SIAM J. Math. Anal.} {\bfseries 38} (2006) 574–594}.

\bibitem{2008MNRAS.383..539M}
D.~{Marinucci}, D.~{Pietrobon}, A.~{Balbi}, P.~{Baldi}, P.~{Cabella}, G.~{Kerkyacharian} et~al., \emph{{Spherical needlets for cosmic microwave background data analysis}}, \href{https://doi.org/10.1111/j.1365-2966.2007.12550.x}{\emph{\mnras} {\bfseries 383} (2008) 539} [\href{https://arxiv.org/abs/0707.0844}{{\ttfamily 0707.0844}}].

\bibitem{2008arXiv0811.4440G}
D.~{Geller} and A.~{Mayeli}, \emph{{Continuous Wavelets on Compact Manifolds}}, \href{https://doi.org/10.48550/arXiv.0811.4440}{\emph{arXiv e-prints} (2008) } [\href{https://arxiv.org/abs/0811.4440}{{\ttfamily 0811.4440}}].

\bibitem{HEALpix}
K.M.~{G{\'o}rski}, E.~{Hivon}, A.J.~{Banday}, B.D.~{Wandelt}, F.K.~{Hansen}, M.~{Reinecke} et~al., \emph{{HEALPix: A Framework for High-Resolution Discretization and Fast Analysis of Data Distributed on the Sphere}}, \href{https://doi.org/10.1086/427976}{\emph{\apj} {\bfseries 622} (2005) 759} [\href{https://arxiv.org/abs/astro-ph/0409513}{{\ttfamily astro-ph/0409513}}].

\bibitem{likelihood}
S.~{Hamimeche} and A.~{Lewis}, \emph{{Likelihood analysis of CMB temperature and polarization power spectra}}, \href{https://doi.org/10.1103/PhysRevD.77.103013}{\emph{\prd} {\bfseries 77} (2008) 103013} [\href{https://arxiv.org/abs/0801.0554}{{\ttfamily 0801.0554}}].

\bibitem{primB}
N.~{Katayama} and E.~{Komatsu}, \emph{{Simple Foreground Cleaning Algorithm for Detecting Primordial B-mode Polarization of the Cosmic Microwave Background}}, \href{https://doi.org/10.1088/0004-637X/737/2/78}{\emph{\apj} {\bfseries 737} (2011) 78} [\href{https://arxiv.org/abs/1101.5210}{{\ttfamily 1101.5210}}].

\bibitem{Mangilli_2021}
A.~{Mangilli}, J.~{Aumont}, A.~{Rotti}, F.~{Boulanger}, J.~{Chluba}, T.~{Ghosh} et~al., \emph{{Dust moments: towards a new modeling of the galactic dust emission for CMB B-modes analysis}}, \href{https://doi.org/10.1051/0004-6361/201937367}{\emph{\aap} {\bfseries 647} (2021) A52} [\href{https://arxiv.org/abs/1912.09567}{{\ttfamily 1912.09567}}].

\bibitem{Ritacco_2023}
A.~{Ritacco}, F.~{Boulanger}, V.~{Guillet}, J.-M.~{Delouis}, J.-L.~{Puget}, J.~{Aumont} et~al., \emph{{Dust polarization spectral dependence from Planck HFI data. Turning point for cosmic microwave background polarization-foreground modeling}}, \href{https://doi.org/10.1051/0004-6361/202244269}{\emph{\aap} {\bfseries 670} (2023) A163} [\href{https://arxiv.org/abs/2206.07671}{{\ttfamily 2206.07671}}].

\bibitem{cMILC}
M.~{Remazeilles}, A.~{Rotti} and J.~{Chluba}, \emph{{Peeling off foregrounds with the constrained moment ILC method to unveil primordial CMB B modes}}, \href{https://doi.org/10.1093/mnras/stab648}{\emph{\mnras} {\bfseries 503} (2021) 2478} [\href{https://arxiv.org/abs/2006.08628}{{\ttfamily 2006.08628}}].

\bibitem{ocMILC}
A.~{Carones} and M.~{Remazeilles}, \emph{{Optimization of foreground moment deprojection for semi-blind CMB polarization reconstruction}}, \href{https://doi.org/10.1088/1475-7516/2024/06/018}{\emph{\jcap} {\bfseries 2024} (2024) 018} [\href{https://arxiv.org/abs/2402.17579}{{\ttfamily 2402.17579}}].

\bibitem{Beamsys}
C.~{Leloup}, G.~{Patanchon}, J.~{Errard}, C.~{Franceschet}, J.E.~{Gudmundsson}, S.~{Henrot-Versill{\'e}} et~al., \emph{{Impact of beam far side-lobe knowledge in the presence of foregrounds for LiteBIRD}}, \href{https://doi.org/10.1088/1475-7516/2024/06/011}{\emph{\jcap} {\bfseries 2024} (2024) 011} [\href{https://arxiv.org/abs/2312.09001}{{\ttfamily 2312.09001}}].

\bibitem{HWPsys}
M.~{Monelli}, E.~{Komatsu}, T.~{Ghigna}, T.~{Matsumura}, G.~{Pisano} and R.~{Takaku}, \emph{{Impact of half-wave plate systematics on the measurement of CMB B-mode polarization}}, \href{https://doi.org/10.1088/1475-7516/2024/05/018}{\emph{\jcap} {\bfseries 2024} (2024) 018} [\href{https://arxiv.org/abs/2311.07999}{{\ttfamily 2311.07999}}].

\end{thebibliography}\endgroup

\end{document}